\documentclass[lettersize,journal]{IEEEtran}
\usepackage{amsmath,amsfonts}
\usepackage{algorithmic}
\usepackage{algorithm}
\usepackage{array}

\usepackage{textcomp}
\usepackage{stfloats}
\usepackage{url}
\usepackage{verbatim}
\usepackage{graphicx}
\usepackage{cite}
\usepackage{booktabs}

\usepackage{subcaption}
\usepackage{xcolor}

\newcommand{\brayan}[1]{{\color{black}{#1}}}

\newcommand{\bacca}[1]{{\color{black}{#1}}}

\newcommand{\bb}[1]{\textbf{#1}}
\newcommand{\ul}[1]{\underline{#1}}

\usepackage{hyperref}
\hypersetup{
    colorlinks=true,
    linkcolor=blue,
    filecolor=magenta,      
    urlcolor=blue,
    citecolor = red,
}

\begin{document}

\title{Deep Scene-Driven Ordering of Hadamard Basis \\ for Single-Pixel Spectral Imaging}


\author{ Brayan Monroy, Hans Garcia, Henry Arguello,
\bacca{Jorge Bacca}  \vspace{-0.4em}

\thanks{Brayan Monroy, Hans Garcia, Henry Arguello and Jorge Bacca are with the Universidad Industrial de Santander, Bucaramanga, 680001, Colombia. }

}

\maketitle

\begin{abstract}
Spectral images are highly valuable for various applications, including environmental monitoring and precision agriculture. However, the high cost of specialized sensors limits the wide use of this technology in numerous applications. Current alternatives to acquire high spatial-spectral resolution spectral images, like Single-Pixel Imaging (SPI) enhanced with Deep Optical Coding Design (DOCD), have limitations due to their non-feedback optical designs, leading to limited image quality, with optimal performance achieved only for the specific scenes used during training. This work reformulates the DOCD framework to handle the scene-driven ordering of the Hadamard basis within the SPI architecture for spectral imaging. Taking into account that SPI usually acquires hundreds of snapshots, our approach introduces a scene-driven ordering of the Hadamard matrix for flexible SPI modulation pattern selection based on scene characteristics in an end-to-end optimization. Simulations on spectral datasets and real test-bed acquisitions demonstrate the effectiveness of the proposed method in improving the quality of VIS and NIR spectral images compared to fixed designs. 
\end{abstract} 

\begin{IEEEkeywords}
Hadamard single-pixel, spectral imaging, scene-driven sensing, deep learning, near-infrared spectrum.
\end{IEEEkeywords}

\section{Introduction}

Spectral images encode spatial information across multiple electromagnetic wavelengths, enabling analysis of material-specific absorption and reflectance beyond conventional RGB imaging~\cite{garini2006spectral, bacca2023computational}. The spectral behavior of materials and objects, or spectral signature, is highly informative for detection, classification, and segmentation in applications such as remote sensing, medical imaging, and precision agriculture~\cite{shaw2003spectral}. Across the ultraviolet, visible, and infrared ranges, spectral imaging reveals properties not available from single-intensity measurements; for example, thermal radiation from objects near room temperature is prominent in the infrared~\cite{teena2014thermal}, and vegetation indices like NDVI benefit from non-visible wavelengths~\cite{zhu2018review}. However, challenges remain in acquiring high-resolution spectral images across spectral ranges due to sensor technology and cost constraints, which limits commercial accessibility~\cite{rogalski2022scaling, rogalski2016challenges}. In addition, many spectral cameras rely on scanning or multiplexing strategies that require a large number of measurements, leading to long acquisition times and restricting the use of spectral imaging in dynamic scenarios.

The Single-Pixel Imaging (SPI) system has emerged as a promising and cost-effective solution to acquire spectral images, leveraging the principles of Compressive Sensing (CS)~\cite{duarte2008single}. In contrast to conventional scanning methods, which sequentially capture subsets of spectral images, SPI systems capture inner products between the entire scene and a set of modulation patterns, substantially reducing hardware complexity and cost. Specifically, in the SPI system, the spatial resolution is defined by modulation patterns, such as coded apertures implemented through a Digital Mirror Device (DMD)~\cite{sampsell1994digital}, and the resolution of the spectrometer defines the spectral resolution. This split between spatial and spectral resolutions offers a cost-effective approach, particularly beneficial in spectral ranges where conventional 2D sensors face technological and physical limitations~\cite{centrone2015infrared, gibson2020single}.

However, a well-known limitation of SPI systems is the need to use many modulation patterns, resulting in long acquisition times~\cite{garcia2020optimized,monroy2023deep,bacca2020coupled}. Consequently, significant efforts have been made to reduce the required snapshots to speed up image acquisition and recovery processes. In particular, Hadamard SPI (HSPI) introduces an approach where modulation patterns are derived from rows of the orthogonal Hadamard matrix, which are subsequently reshaped into two-dimensional matrices~\cite{vaz2020image}. The binary and orthogonal properties of the Hadamard matrix enable practical optical implementations and fast image recovery via matrix multiplication~\cite{zhang2017hadamard}. In compression, only a subset of its rows is used, reducing the number of modulation patterns and thus the acquisition time. Several ordering strategies have been proposed to select Hadamard patterns that best preserve image quality~\cite{vaz2020image}. These orderings are based on the number of sign changes~\cite{agaian2011hadamard}, the count of blocks of Hadamard patterns~\cite{yu2019super, sun2017russian}, the maximization of total variation~\cite{yu2020super}, and the use of the structural geometry of the sensing path~\cite{lopez2022efficient, cai2022detail}.

\begin{figure*}[!t]
    \centering
    \includegraphics[width=\linewidth]{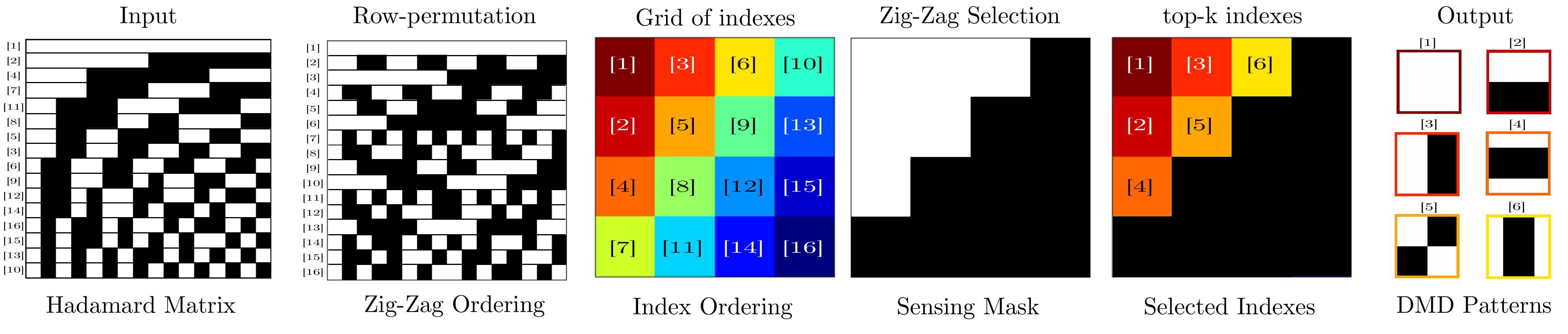}\vspace{-1em}
    \caption{Generation of DMD Hadamard patterns. Specifically, we select The zig-zag ordering, constructed by permuting the rows of the Hadamard matrix of size $n$. Subsequently, indexes of each row of the Hadamard matrix are rearranged into a 2D grid of indexes of size $\sqrt{n}\times \sqrt{n}$. The sensing masks are constructed by selecting the top $k$ patterns based on the initial index values. Finally, the selected DMD patterns correspond to the two-dimensional reshaping of the selected rows.}
    \label{fig:hadamardgen} \vspace{-1em}
\end{figure*}

On the other hand, integrating deep learning techniques with compressive sensing has led to significant advancements in spectral imaging~\cite{huang2022spectral}. In particular, deep neural networks have improved the reconstruction of spectral images in compressive spectral imaging by learning nonlinear data representations~\cite{choi2017} or through their incorporation into optimization recovery algorithms~\cite{zheng2021deep, wang2019hyperspectral}. In addition, deep learning architectures have been used to improve compressive spectral imaging systems directly related to acquisition by designing optical coding elements using the well-known deep optical coding design (DOCD) scene-driven framework. Specifically, optical elements are treated as adjustable parameters in the design of acquisition systems, allowing joint optimization of hardware and software components~\cite{arguello2023deep}. The optimization strategies employed by DOCD, aimed at maximizing data quality, contribute to improvements across a diverse range of computational imaging tasks, including compressed spectral imaging~\cite{bacca2021deep}, privacy-preserving pose estimation~\cite{hinojosa2022privhar}, and privacy-scene captioning~\cite{arguello2022optics}.

Hadamard ordering and the DOCD approaches have practical limitations when applied to different scenes. \brayan{Hadamard ordering assumes fixed order sequence for all scenes, overlooking the unique attributes of each spectral image}, thereby losing specific features of individual scenes in the recovery process~\cite{monroy2023deep}. Similarly, DOCD converges towards a non-feedback design once the parameters are trained with a specific dataset. In this sense, a scene-driven acquisition protocol informed by previous captures can potentially improve image reconstruction by selecting subsequent sensing masks that consider the specific scene characteristics.

In this work, we propose a computational framework that integrates a two-stage sensing strategy with a learned ordering of the Hadamard basis for single-pixel spectral imaging. 
\brayan{Our main contribution is a scene-driven mechanism that, given a small set of low-frequency Hadamard measurements acquired at the beginning of each experiment, predicts a binary support over the remaining Hadamard coefficients and thereby induces a scene-dependent ordering of the Hadamard basis.} \brayan{This approach differs from classical adaptive sensing schemes that rely on hand-crafted decision rules or multi-step feedback policies, as well as from methods that learn free-form sensing matrices: in our case, the sensing patterns remain strict Hadamard $\{\pm 1\}$ patterns, fully compatible with DMD-based hardware, and the learning module only reorders which of these patterns are measured next.} \brayan{Moreover, the proposed ordering is reconstruction-agnostic, allowing us to plug it into different inverse solvers without retraining, and we show experimentally that a model trained on a multispectral satellite dataset can be transferred, without fine-tuning, to a real NIR single-pixel imaging setup, where it consistently outperforms traditional fixed Hadamard orderings.}

\begin{figure}[!t]
    \centering
    \includegraphics[width=\linewidth]{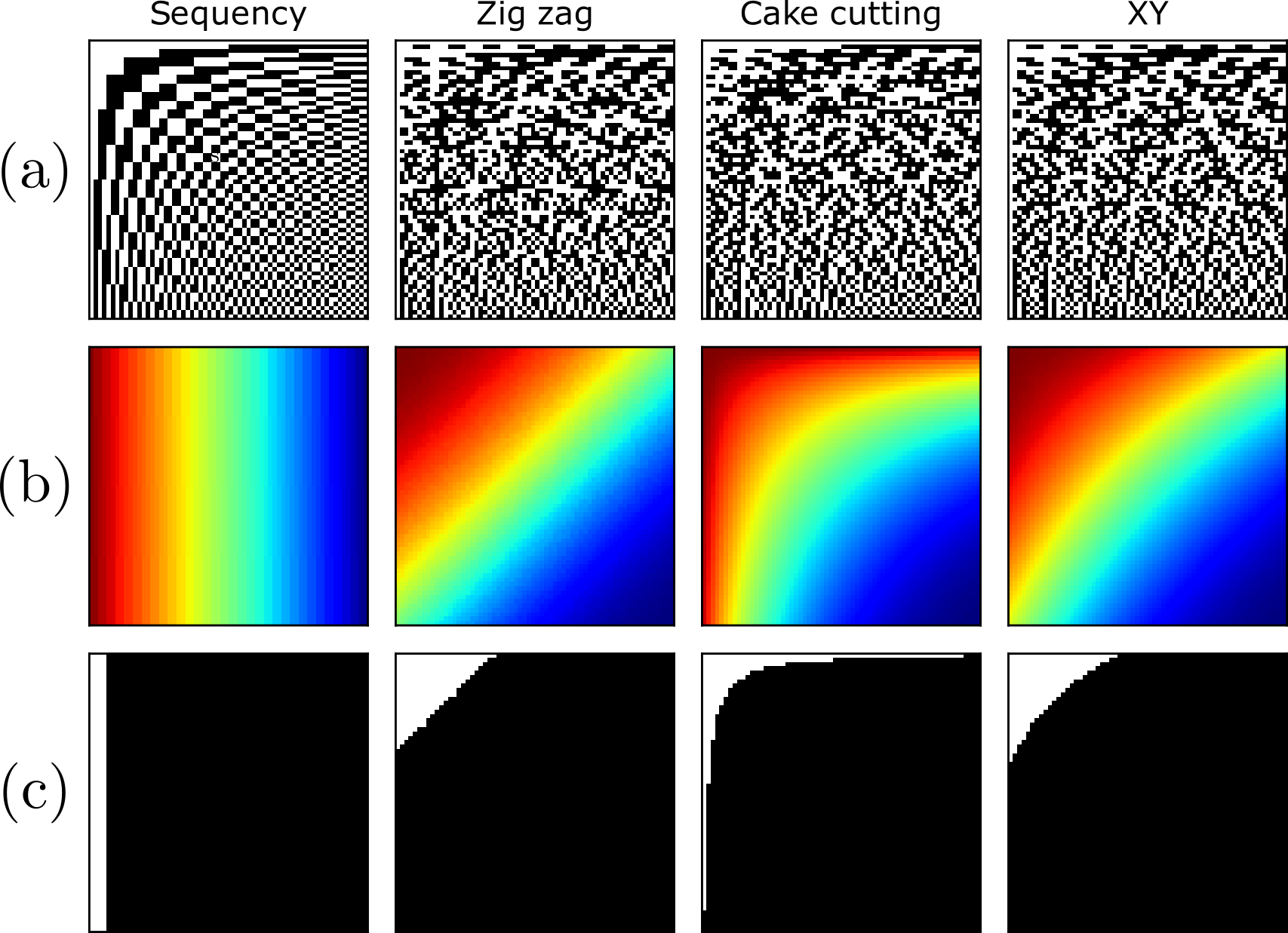} \vspace{-1.3em}
    \caption{ Hadamard Ordering Strategies. (a) The first row illustrates state-of-the-art orderings of the Hadamard matrix. (b) The second row depicts the geometry of coefficients relevant to the sensing path of the 2D Hadamard spectrum. (c) The final row displays the binary masks for selecting the X\% of 2D Hadamard spectrum to be sensed, where white stands for acquired coefficients and black for non-acquired.}   \vspace{-1.3em}
    \label{fig:ordering}
\end{figure}

\section{Related Work}

\subsection{Hadamard Ordering}

In the HSPI system, the entire acquisition scenario consists of employing all rows of the Hadamard matrix as modulation patterns during the acquisition process; that is, acquire $n$ snapshots for an image with $\sqrt{n}\times\sqrt{n}$ pixels, as presented in Figure~\ref{fig:hadamardgen}. Consequently, the entire acquisition is time-consuming, and most of the Hadamard coefficients of a natural image are close to zero, indicating a sparse representation. This approach relies on many unnecessary snapshots~\cite{duarte2008single}. As an alternative, the compression scenario consists of properly selecting some rows of the Hadamard matrix to capture the most relevant coefficients directly in the sensing~\cite{vaz2020image}. Following this strategy, HSPI ordering methods have been proposed to address row selection, where the rows of the Hadamard matrix are rearranged based on the relevance of each row for spatial image representation before sensing, as illustrated in Figure~\ref{fig:ordering}. Then, the $k$-top rows based on this ordered Hadamard matrix are selected, assuming they are associated with the most relevant coefficients that successfully represent the spatial information in the image.

Among HSPI ordering methodologies, \textbf{Zig Zag} technique introduced in \cite{lopez2022efficient}, combines the benefits of sequence ordering and zigzag traversal of patterns. Conversely, the \textbf{Cake-Cutting} approach \cite{yu2019super} strategically rearranges the internal block count in modulation patterns in ascending order. In a different strategy, the \textbf{XY} ordering~\cite{cai2023detail} generates patterns grounded in Cartesian coordinates. Lastly, the \textbf{Russian Dolls} strategy \cite{sun2017russian} is based on optimized pattern ordering, exploiting the symmetry of the Hadamard matrix. Ordering approaches enable the progressive acquisition of coarse to fine details in spectral image reconstruction. These orderings are based on observations of the general behavior of the images to predetermined sets of patterns before sensing, which could lead to suboptimal solutions for spectral images that substantially differ in their Hadamard coefficient distribution from the predefined ordering approach.

\subsection{End-to-End Deep Optical Coding Design}
\brayan{The DOCD methodology consists of designing optical elements in computational imaging systems by jointly optimizing optics and neural network parameters in a wide-spread end-to-end (E2E) training scheme~\cite{arguello2023deep}.} The DOCD methodology allows for the design of optical elements for specific tasks, such as spectral image restoration~\cite{arguello2021shift} or high-level tasks, such as pose estimation~\cite{hinojosa2022privhar}. Several notable works employ the DOCD framework. For example, the work in \cite{chakrabarti2016learning} learns the color multiplexing pattern of the camera sensor by encoding it as a sensor layer and jointly training it with the reconstruction network; the work in \cite{bacca2020coupled} optimizes binary modulation patterns for compressive image classification and reconstruction; the work in \cite{arguello2021shift} focus on the deep optical design of a shift-variant diffractive optical element for the spectral imaging system, and the design in \cite{bacca2021deep} incorporates proper regularization of optical trainable parameters to achieve effective and efficient optical designs. However, once the optical parameters have been adjusted in training, the learned optics remain fixed during subsequent captures. This inflexibility can limit the performance of DOCD-based systems when utilized with scenes significantly different from those found during training.

\subsection{Adaptive Compressive Sampling}

Several alternatives have been proposed to tailor a different modulation pattern for each sensed image. For example, the authors in \cite{diaz2018adaptive} propose a gradient-thresholding algorithm to compute consecutive color-coded apertures from a low-resolution estimate. Specifically, in the case of SPI, some works propose scene-driven methodologies for selecting coding functions based on wavelet-based scans. For instance, \cite{dekel2008adaptive} introduces a tree-structured gradual selection that considers a "father–son" relationship between wavelet coefficients to predict the relevant coefficients at progressively finer scales. The extended work in \cite{averbuch2012adaptive} modifies the prediction strategy via dictionary modeling, discrete probability estimation, and mutual information to determine whether a coefficient is significant. However, these approaches employ real-valued modulation patterns, which are not trivial to implement in SPI, unlike the binary patterns used with the Hadamard matrix.

For HSPI, scene-guided strategies leverage side information from auxiliary cameras to optimize the selection of modulation patterns. For example, \cite{radwell2014single} uses measurements from a previous frame to decide which patterns to select by sorting those measurements by magnitude. From a spatial decimation perspective, superpixel maps have been used to preserve structural information during pattern selection, where the superpixel map is constructed via SLIC \cite{garcia2020optimized} or estimated by a neural network \cite{monroy2023deep}. However, these side-information strategies require two acquisition systems, one for side-information and one for HSPI, introducing spatial misalignment and varying sensor responses that complicate practical deployment.  In contrast, we perform online, content-aware selection of modulation patterns by exploiting the previously captured measurements from the same HSPI system. \brayan{While this concept showed promising results in our preliminary study \cite{monroy2024predicting}, this work extends it by introducing a reconstruction-agnostic, scene-driven basis ordering framework. We mathematically formulate the selection masks, evaluate them across multiple spectral datasets using advanced Plug-and-Play and regularization-based solvers, and validate robustness and hardware compatibility on a NIR spectral imaging testbed under out-of-distribution conditions.}

\begin{figure}[b]
    \centering
    \includegraphics[width=\linewidth]{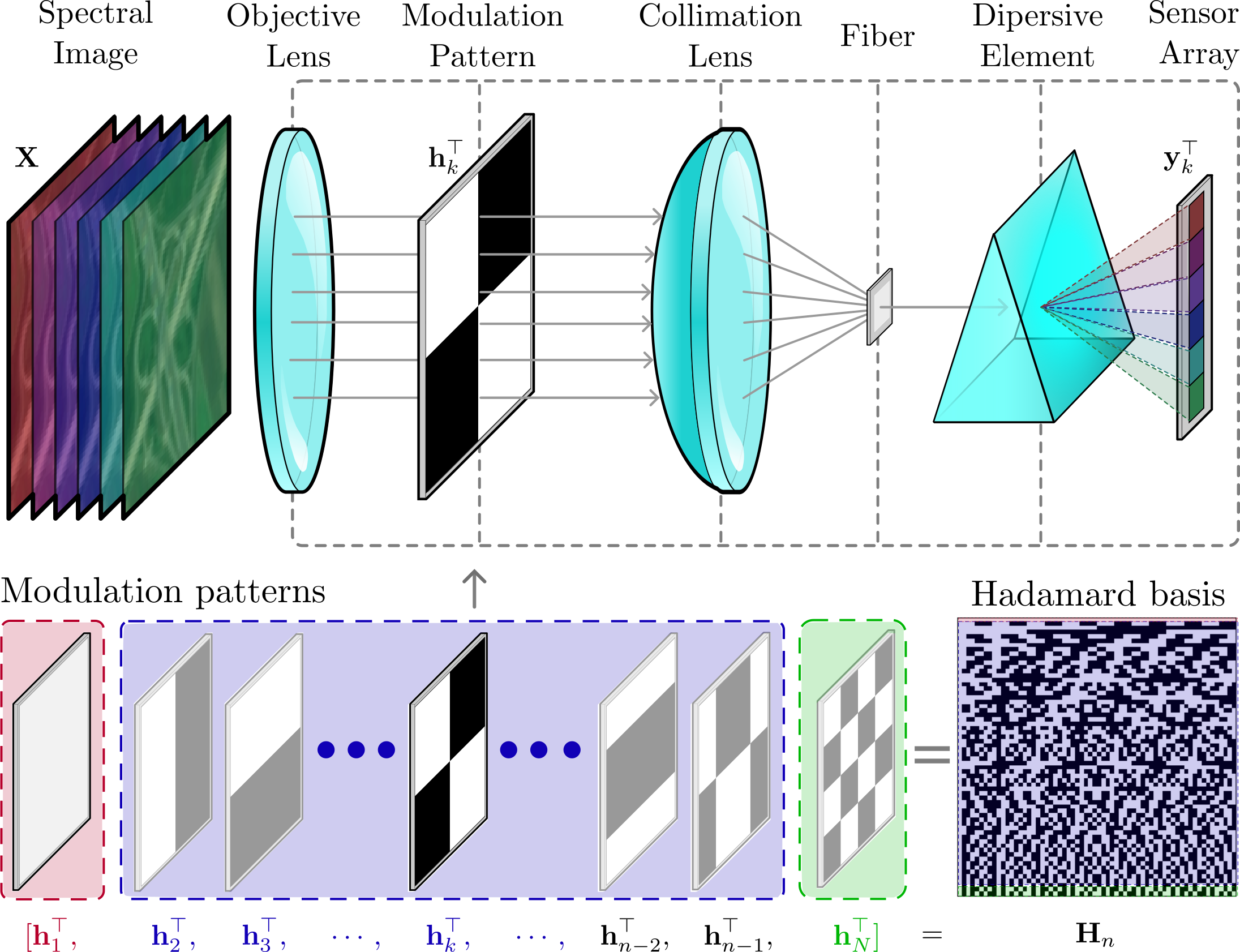}
    \caption{Hadamard Single Pixel Imaging. The starting spectral image is spatially modulated using a specific pattern and then focused through a collimating lens, after which the spectral information is detected and recorded by a sensor array.}
    \label{fig:hsi} \vspace{-1.5em}
\end{figure}

\section{Hadamard Single-Pixel Imaging}

In the SPI system, a spectrometer captures the inner product between a set of modulation patterns $\{ \mathbf{h}_{i} \}_{i=1}^n$ and the spatial vectorization of a NIR spectral image, as presented in Figure~\ref{fig:hsi}. The NIR spectral image, denoted as $\mathbf{X} \in \mathbb{R}^{n \times c}$, is a matrix where $n=hw$ represents the total spatial pixels, with $h$ and $w$ as spatial dimensions, and $c$ denotes the number of spectral bands. Here, $\mathbf{X}$ is the horizontal stack of vectorized spatial information by bands.
In the Hadamard case, the modulation patterns consist of rows of a Hadamard matrix of $n$ order, denoted as $\mathbf{H} \in \{ -1, 1 \}^{n \times n}$, with each row of $\mathbf{H}$ representing a modulation pattern $\mathbf{h}_i$ of the set $\{ \mathbf{h}_i \}_{i=1}^n$. Consequently, the acquisition process of the HSPI system involves the linear transformation of the NIR spectral image using the orthogonal Hadamard basis. This transformation is equivalent to acquiring the Hadamard coefficients $\mathbf{Y} \in \mathbb{R}^{n\times c}$ for each spectral band of the NIR image, which can be defined as follows:
\begin{equation} \mathbf{Y} = \mathbf{H} \mathbf{X} + \mathbf{Z},
\end{equation}
where, $\mathbf{Z}$ denotes the noise. In this sense, using the complete set of modulation patterns corresponds to a full acquisition of the spectral image in the Hadamard spectrum, which means that no compression occurs. However, images in the Hadamard spectrum often exhibit a sparse representation, characterized by a high number of near-zero coefficients, which implies an unnecessary amount of modulation patterns. Therefore, the compression scenario involves selecting a subset of the rows of the Hadamard matrix to preserve the most relevant information about the NIR spectral image and reduce the number of modulation patterns. Mathematically, the subsampling of the Hadamard matrix can be defined as follows: \begin{equation}
\mathbf{Y} = \mathbf{M} \mathbf{H} \mathbf{X},
\end{equation}
here, $\mathbf{M} = \text{diag}(\mathbf{m})$ is a diagonal matrix that selects a set of Hadamard rows based on the entries of the binary selection mask $\mathbf{m} \in \{0, 1\}^{n}$. The sampling ratio is given by $\delta = \Vert\mathbf{m}\Vert_0 / n$. Once captured, a coarse spectral image reconstruction based on the solution of $||\mathbf{Y} - \mathbf{M}\mathbf{H} \mathbf{X}||_F$ can be obtained using the orthogonal characteristics of the Hadamard matrix, which acts as the transpose of the sensing matrix as
\begin{equation}
\hat{\mathbf{X}} = \frac{1}{n} \mathbf{H}^{\top} \mathbf{Y}.\label{eqn:recovery}
\end{equation} 
This is based on the fact that the $\textbf{Y}$ maintains the dimensionality with zeros in the non-acquired elements.

\section{Method}

The proposed scene-driven Hadamard ordering divides the HSPI system into two sequential sensing stages: (i) predetermined sensing and (ii) scene-driven sensing. In the first stage, we acquire an initial subset of the Hadamard spectrum using a predefined sensing path from state-of-the-art Hadamard ordering algorithms. In the second stage, we feed these initial measurements to a deep neural network, which predicts the ordering of the remaining coefficients to select a second subset of modulation patterns in an end-to-end manner. The next subsection presents the acquisition protocol, neural network architecture, and training procedure.

\subsection{Acquisition protocol}

\begin{figure*}[!t]
    \centering
    \includegraphics[width=0.95\linewidth]{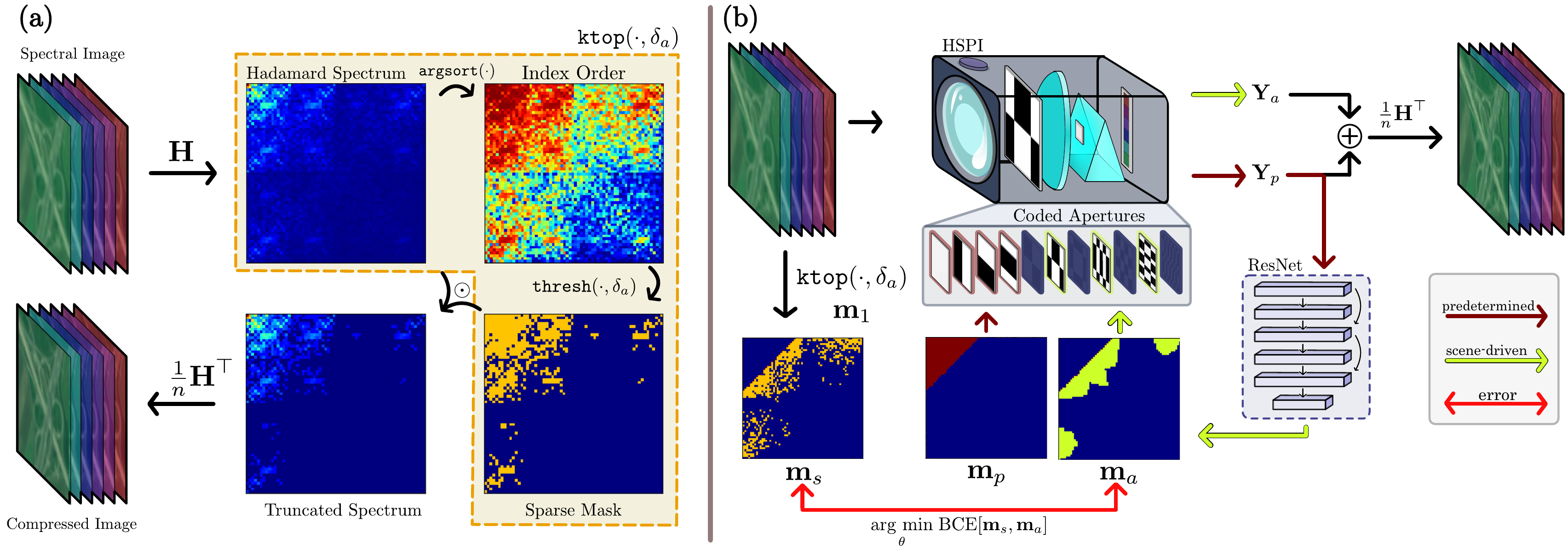} \vspace{-1em}
    \caption{scene-driven Hadamard Single Pixel Imaging. (a) Construction of reference sensing mask $\textbf{m}_s$: the reference sensing mask $\textbf{m}_s$ are constructed by selection of the $k_a$ top coefficients in the Hadamard spectrum. (b) Proposed acquisition protocol, a predetermined sensing acquisition $\textbf{Y}_p$ based on sensing mask $\textbf{m}_p$ guides the selection of scene-driven sensing mask~$\textbf{m}_a$.}
    \label{fig:acquistion} \vspace{-1em}
\end{figure*}

The acquisition protocol involves obtaining the Hadamard coefficients from the scene-driven row selection matrix $\mathbf{M} = \text{diag}(\mathbf{m}_p + \mathbf{m}_a)$, where $\mathbf{m}_p$ represents the predetermined selection rows and $\mathbf{m}_a$ corresponds to the rows selected in a scene-driven manner, as illustrated in Figure~\ref{fig:acquistion}. The total sampling ratio is defined as $\delta = \delta_p+\delta_a$, where $\delta_p=k_p/n$ and $\delta_a=k_a/n$ are the predetermined and scene-driven sampling ratios, with $k_p=\Vert \mathbf{m}_p\Vert_0$ and $k_a=\Vert\mathbf{m}_a\Vert_0$ denoting the corresponding numbers of selected rows, respectively.

\textbf{Predetermined sensing.} The predetermined sensing step consists of acquiring a subset of the Hadamard matrix using a consistent approach across all images, following the current state-of-the-art sensing method (typically Zig-Zag~\cite{lopez2022efficient}). It is worth noting that we assume the statistical differences in the ordering of coefficients are negligible at this stage. Mathematically, this process can be represented as
\begin{equation}
\mathbf{Y}_p = \mathbf{M}_p \mathbf{H} \mathbf{X},
\end{equation}
where $\mathbf{M}_p = \text{diag}(\mathbf{m}_p)$ is the predetermined selection matrix, which selects $k_p$ rows of the Hadamard matrix based on a predefined Hadamard ordering algorithm.

\vspace{1em}
\textbf{Scene-driven Hadamard Ordering.} Following the initial acquisition $\mathbf{Y}_p$, a deep neural model $\mathcal{P}_\theta$ estimates the ordering of the most relevant coefficients in the non-acquired spectrum. Mathematically, this estimation is represented as
\begin{equation}
\mathbf{m}_a = \mathcal{P}_{\theta}(\mathbf{Y}_p),
\end{equation}
\brayan{Here, $\mathbf{m}_a \in \{0,1\}^n$ is the postprocessed binary selection mask. During training, the parameters $\theta$ of $\mathcal{P}_\theta$ are optimized so that $\mathbf{m}_a$ matches a reference support $\mathbf{m}_s$, defined by the top-$k_a$ Hadamard coefficients of each training sample, as detailed in the E2E training procedure below. In practice, $\mathcal{P}_\theta$ first outputs sigmoid scores in $[0,1]^n$; at test time, entries already selected by $\mathbf{m}_p$ are masked out, and the $k_a$ largest remaining scores are set to one to form $\mathbf{m}_a$, with all others set to zero. The trained network is then fixed and, for each new scene, $\mathbf{m}_a$ defines the scene-driven selection matrix $\mathbf{M}_a = \text{diag}(\mathbf{m}_a)$. An entry $m_a(i)=1$ means the $i$-th row of the Hadamard matrix is selected and, in the second acquisition stage, implemented on the DMD as a pair of complementary $\{0,1\}$ patterns following the optical protocol. Since all entries selected by $\mathbf{m}_a$ are acquired within one second-stage block, their internal order does not affect reconstruction; “ordering” instead denotes the priority/support of coefficients under a fixed budget. This procedure induces the second Hadamard coefficient ordering from the first acquisition $\mathbf{Y}_p$.
}

\textbf{Scene-driven Sensing.} The scene-driven sensing consists of acquiring the remaining coefficients based on the previously estimated ordering. This acquisition process is formulated as
\begin{equation}
\mathbf{Y}_a = \mathbf{M}_a \mathbf{H} \mathbf{X}.
\end{equation}
Subsequently, as both predetermined and scene-driven acquisitions are associated with the same Hadamard basis, and the captured coefficients are mutually exclusive, the coarse spectral reconstruction can be carried out from both acquisitions by solving the following optimization problem
\begin{equation}
\hat{\mathbf{X}} \in \underset{\textbf{X}}{\text{arg min}} \quad f(\textbf{X};\textbf{Y},\textbf{H}) + \lambda g(\textbf{X}),
\end{equation}
with $\textbf{Y}=\textbf{Y}_p+\textbf{Y}_a$. Here, $f(\cdot)$ is a measurement consistency term and $g(\cdot)$ is the signal regularization term. When $f(\cdot)$ is the $\ell_2$ norm, that is, $f(\textbf{X};\textbf{Y},\textbf{H})= \Vert \textbf{Y} - \textbf{MH} \textbf{X} \Vert_2^2$ and $\lambda = 0$, the closed-form solution is given by Equation~\eqref{eqn:recovery}.

\subsection{E2E Training procedure}

To address the optimization of the scene-driven estimation model, denoted as $\mathcal{P}_\theta$, we employ the E2E framework. This framework encompasses the joint modeling of the two sensing acquisitions and the reconstruction of spectral images within a network architecture, illustrated in Figure~\ref{fig:acquistion}. Our final goal is to optimize the trainable parameters, represented by $\theta$, within the scene-driven estimation model $\mathcal{P}_\theta$. This optimization process aims to minimize a defined cost function in the training data set, denoted $\mathcal{L}(\cdot)$. This training optimization problem can be formally expressed as follows
\begin{equation}
\theta^* \in \underset{\theta}{\text{arg min}} \quad \mathbb{E}_{\textbf{X}} [ \mathcal{L}( \textbf{X}, \hat{\textbf{X}}; \mathcal{P}_\theta ) ]. \label{eqn:training}
\end{equation}
where $\mathbb{E}_{\textbf{X}}$ represents the expected value of a training dataset sampled randomly in each epoch following a stochastic gradient descent approach to solving Equation~\eqref{eqn:training}.

\textbf{Scene-driven Sensing as Binary Classification.} To address the selection of modulation patterns with the scene-driven estimation model, denoted as $\mathcal{P}_\theta$, we model the problem of selection of most relevant coefficients as a binary classification problem. Since each spectral image can be represented in the Hadamard spectrum from its linear combination of vector basis $\textbf{h}_k$ and intensities $\textbf{y}_k$ as follows
\begin{equation}
     \textbf{X} = \sum_{k=1}^{n}  \textbf{h}_k \otimes \textbf{y}_k.
\end{equation}
We propose splitting the Hadamard coefficients into three classes: i) the predetermined coefficients acquired in the first step of the scheme, ii) the top $k_a$ Hadamard coefficients by magnitude, which serve as the ground-truth labels learned by the scene-driven selection model, and iii) the least relevant coefficients. \brayan{The identification of these informative components is derived from the full Hadamard decomposition of each training sample, where coefficients are ranked by absolute value. The highest-magnitude coefficients concentrate the dominant portion of the scene energy and therefore impose the greatest influence on the reconstruction.} Correspondingly, the modulation patterns are divided into $\{ \hat{\textbf{h}}_i \}_{i=1}^{k_p}$ for the predetermined patterns, $\{ \hat{\textbf{h}}_j \}_{j=k_p+1}^{k_{a}+k_{p}}$ for the scene-driven patterns, and $\{ \hat{\textbf{h}}_k \}_{k=k_a+k_p+1}^{n}$ for the unused patterns, allowing the spectral image to be represented as follows
\begin{equation}
    \textbf{X} = \underbrace{ \sum_{i=1}^{k_p}   \hat{\textbf{h}}_i \otimes \textbf{y}_i }_{predetermined} 
    + \underbrace{ \sum_{j=k_p+1}^{k_{a}+k_{p}}   \hat{\textbf{h}}_j \otimes \textbf{y}_j }_{most \; relevant} 
    +  \underbrace{ \sum_{k=k_a+k_p+1}^{n}   \hat{\textbf{h}}_k \otimes \textbf{y}_k }_{less\; relevant}.
\end{equation}
Let $|\textbf{Y}|_{\Sigma}\in\mathbb{R}^{n}$ denote the spectral-band aggregated Hadamard magnitude, with $|\textbf{Y}|_{\Sigma}(i)=\sum_{\ell=1}^{c}|Y_{i,\ell}|$. Hence, the binary classification mask can be computed as a position-threshold in the ordered Hadamard spectrum $\textbf{m}_s = \texttt{ktop}( \textbf{M}_p^r|\textbf{Y}|_{\Sigma}, k_a)$, discarding the already sensed coefficients in the predetermined sensing by the reverse mask $\textbf{M}_p^r$, as presented in Figure~\ref{fig:acquistion}(b). Finally, the reconstruction optimization problem in Equation~\eqref{eqn:training} can reformulated as a binary classification problem where the deep model $\mathcal{P}_\theta(\cdot)$ estimates the scene-driven sensing mask $\textbf{m}_a$ from a predetermined acquisition $\textbf{Y}_p$ which are minimized with a binary cross-entropy loss function as follows
\begin{equation}
 \theta^* \in \underset{\theta}{\text{arg min}} \; \mathcal{L}:=\text{BCE}[\textbf{m}_{s}, {\mathcal{P}_{\theta}(\mathbf{Y}_p)}].
\end{equation}

\section{Results and Simulations}

\subsection{Experimental setup}

In our neural network design approach, we use a standard model to estimate the Hadamard coefficients dynamically. This choice emphasizes the importance of our scene-driven methodology while avoiding the computational overhead associated with complex neural networks. Our deep neural network architecture is centered around a ResNet-based backbone, incorporating 5 residual blocks alongside batch normalization for enhanced stability. Beginning with an initial feature size of 64, our architecture progressively expands along the hidden layers. We strategically insert max-pooling operators between the residual blocks to condense spatial dimensions.

After extraction of activation features from the ResNet backbone, they are input into a fully connected head comprising three layers. Within this head, the final transformation is applied to the selection matrix vector $\textbf{m}_a$, where the last activations consist of a sigmoid activation. Throughout the training phase, dropout regularizers are introduced between fully connected layers, with a dropout probability parameter set at $90\%$ to mitigate overfitting risks. The ReLU is used as the nonlinear activation function throughout the neural network. In particular, our estimation model have approximately 5.5 million trainable parameters.

In terms of training configuration, we conducted 500 epochs with an initial learning rate of $0.001$, optimizing the model parameters using the Adam optimizer. Unless otherwise stated, a separate predictor is trained for each target sampling ratio, so that $k_p$ and $k_a$ are fixed during training and inference. We used the EuroSAT data set, partitioning it into a train-test split of $0.8/0.2$ for model evaluation. The hyperspectral EuroSAT dataset comprises 27k Sentinel-2 satellite images with a spatial resolution of $64 \times 64$, covering 13 spectral bands spanning the spectrum's visible, near-infrared, and short-wave infrared segments. For ARAD evaluations, the same training protocol is applied on ARAD for each reported sampling ratio, while the real NIR testbed uses the EuroSAT-trained selector at the corresponding sampling ratio without fine-tuning. Pytorch implementation is available on GitHub.

\begin{figure*}[!t]
    \centering
    \includegraphics[width=\linewidth]{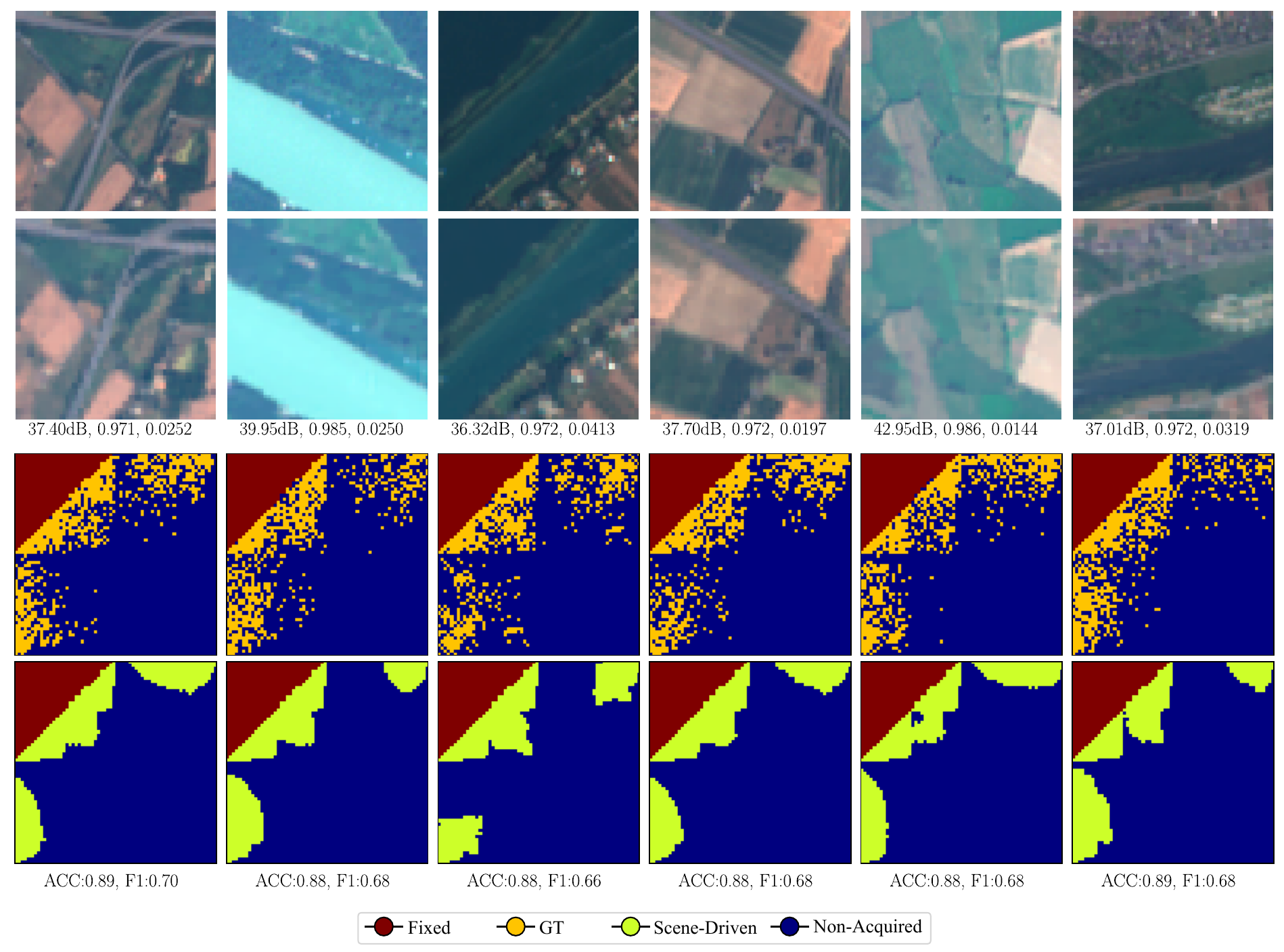} \vspace{-2em}
    \caption{ Image reconstruction from our proposed sensing methodology. a) presents reference images and reconstructed images in the natural domain while b) presents the reference sparse mask $\textbf{m}_s$ and estimated coefficients $\textbf{m}_a$. }
    \label{fig:adaptive2} \vspace{-0.5em}
\end{figure*}

\subsection{Deep scene-driven Ordering Evaluation}

Figure~\ref{fig:adaptive2} provides an evaluation of the proposed scene-driven ordering strategy under a sampling rate of \(\delta=30\%\), meaning only 30\% of the Hadamard spectrum is acquired. The top two rows in Figure~\ref{fig:adaptive2}(a) show the visual comparison between ground-truth and reconstructed RGB composites. The reconstructions exhibit high visual fidelity, with peak PSNR values reaching up to 42.95\,dB and SSIM values up to 0.986. These results validate the effectiveness of the scene-driven selection in preserving spatial detail and spectral integrity under strong measurement constraints. SAM values remain consistently low, with representative examples such as 0.0144 and 0.0197, confirming accurate spectral reconstruction.

To further evaluate the sampling behavior, Figure~\ref{fig:adaptive2}(b) displays the sampling masks over the Hadamard spectrum. For each case, the fixed deterministic measurements \(\mathbf{m}_p\), reference support \(\mathbf{m}_s\), and non-acquired coefficients are visualized in red, orange, and blue, respectively. The scene-driven selected coefficients \(\mathbf{m}_a\) are shown in yellow. The spatial structure of \(\mathbf{m}_a\) adapts in accordance with the underlying signal content, confirming the scene-driven nature of the proposed acquisition policy.

Finally, the bottom row reports quantitative agreement between the learned scene-driven mask \(\mathbf{m}_a\) and the ground-truth informative support \(\mathbf{m}_s\) in terms of Accuracy (ACC) and F1-score. Across all scenes, ACC remains stable around 0.88–0.89, while the F1-score hovers around 0.68, suggesting that although the estimation model does not perfectly match the reference, it provides sufficiently informative samples to enable high-quality image reconstruction. These results demonstrate the potential of the proposed method, which could be further improved with enhanced supervision or more sophisticated architectures.

\begin{figure*}[!b]
    \centering
    \includegraphics[width=\linewidth]{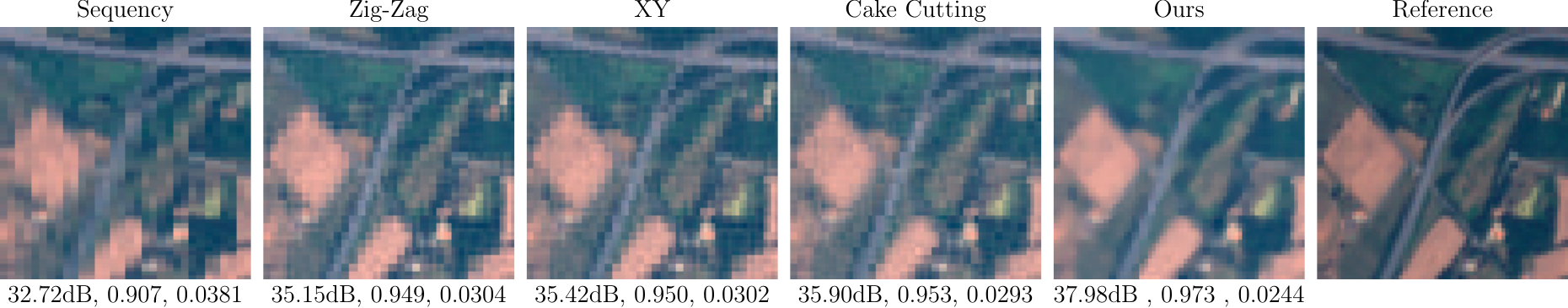}
    \includegraphics[width=\linewidth]{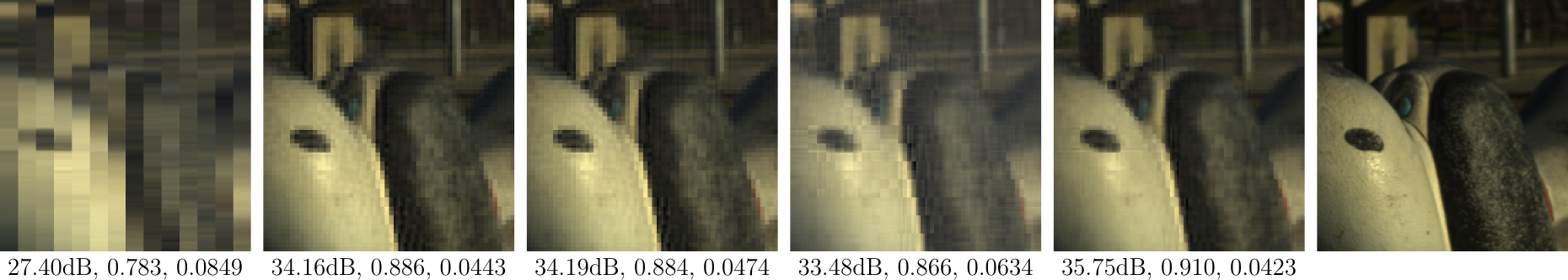}
    \caption{Qualitative comparison between the proposed scene-driven ordering and four fixed Hadamard ordering strategies: Sequency, Zig-Zag~\cite{lopez2022efficient}, XY~\cite{cai2023detail}, and Cake-Cutting~\cite{yu2019super}. The first row correspond to the EUROSAT dataset with sampling rate $\delta = 30\%$, while the second row correspond to the ARAD dataset with $\delta = 5\%$. Each column presents RGB composites and their corresponding quantitative metrics (PSNR, SSIM, SAM).  Visual results demonstrate that our method yields sharper details, higher PSNR and fewer reconstruction artifacts compared to fixed strategies.}
    \label{fig:visualcomparison}
\end{figure*}

\begin{table}[h] \centering

\caption{\textbf{Scene-driven ordering with different recovery algorithms.} best and second best results highlighted in {\textbf{bold}} and \underline{underline} respectively. \label{tab:comparison}} 

\resizebox{0.99\linewidth}{!}{%
\begin{tabular}{l|l||ccc}
\toprule
$\delta$ & Method   & PSNR ($\uparrow$)     & SSIM ($\uparrow$)    & SAM ($\downarrow$)    \\ \hline \hline 
& Transpose                     & {34.98}     & 0.900   & 0.0363   \\ 
& Wavelet Sparsity~\cite{mallat1999wavelet}                 & {35.54}     & 0.921   & 0.0332   \\ 
10$\%$& Median Filter~\cite{chan2016plug}       & \bb{36.47}     & \bb{0.930}   & \underline{0.0324}   \\
& Total Variation~\cite{aggarwal2016hyperspectral}                   & \underline{36.40}     & \underline{0.926}  & \bb{0.0317}   \\
& Consensus~\cite{buzzard2018plug}                       & {36.05}    & 0.924   & 0.0332   \\ 
\midrule 
& Transpose                       & {37.77}     & 0.948  & 0.0276   \\ 
& Wavelet Sparsity~\cite{mallat1999wavelet}                  &  {38.58}     & 0.962   & 0.0254   \\
20$\%$& Median Filter~\cite{chan2016plug}                   & {39.09}     & 0.962  & 0.0253   \\
& Total Variation~\cite{aggarwal2016hyperspectral}                         &  \bb{39.17}     &  \underline{0.962}  &  \bb{0.0241}   \\  
& Consensus~\cite{buzzard2018plug}                       &  \bb{39.17}  &  \bb{0.963}  & \underline{0.0244}   \\ 
\midrule 
& Transpose                        & {39.76}     & 0.966   & 0.0229   \\ 
& Wavelet Sparsity~\cite{mallat1999wavelet}                   & {40.63}     & 0.975   & 0.0212   \\ 
30$\%$& Median Filter~\cite{chan2016plug}                        & 40.03     & 0.969   & 0.0231   \\ 
& Total Variation~\cite{aggarwal2016hyperspectral}               &  \bb{40.67}     &  \bb{0.974}   &  \bb{0.0209}   \\ 
& Consensus~\cite{buzzard2018plug}                      & \underline{40.65}    & \underline{0.973}   & \underline{0.0210}   \\ 
\bottomrule
\end{tabular}%
}
\end{table}

\break

\subsection{Reconstruction algorithm setup.}

\brayan{The proposed method is compatible with both simple inversion via the transpose operator and Plug-and-Play (PnP) algorithms~\cite{venkatakrishnan2013plug} coupled with well-known denoising regularization priors~\cite{elad2023image}. In this work, we evaluate four reconstruction strategies: (i) linear inversion using the transpose operator, (ii) a wavelet-based sparsity prior~\cite{mallat1999wavelet} implemented through iterative shrinkage in the transform domain, (iii) a median-filter prior used as a non-linear denoiser within the PnP iterations\cite{chan2016plug}, and (iv) Total Variation (TV) regularization~\cite{aggarwal2016hyperspectral}. In addition, we validate the Consensus Equilibrium~\cite{buzzard2018plug} (CE) framework, which combines multiple priors by solving for a fixed point where their denoising and data-consistency updates agree, here fusing the Wavelet and TV priors into a single solver. All solvers are implemented with DeepInverse~\cite{tachella2025deepinverse}.}

\brayan{Table~\ref{tab:comparison} summarizes the reconstruction performance of transpose backprojection, Plug-and-Play priors, and the Consensus Equilibrium (CE) solver under sampling ratios \(\delta=\{10\%,20\%,30\%\}\). As expected, the naive transpose operator yields the lowest PSNR values (34.98dB, 37.77dB and 39.76dB), accompanied by degraded SSIM and elevated SAM metrics, reflecting poor noise suppression. Introducing a wavelet prior provides moderate but consistent gains, peaking at 40.63dB at \(\delta=30\%\). Median filtering proves especially effective at the sparsest sampling (\(\delta=10\%\)), achieving PSNR of 36.47dB and SSIM of 0.930. Total Variation regularization attains the highest PSNR at \(\delta=20\%\) and 30\% (39.17dB and 40.67dB) along with the lowest SAM, indicating strong structural preservation. The CE reconstructor, which fuses wavelet and TV priors, delivers robust PSNRs of 36.05dB, 39.17dB and 40.65dB, consistently high SSIM (up to 0.973) and low SAM (down to 0.0210) across all regimes.  Its stability and near-optimal trade-off between spatial sharpness and spectral fidelity justify its adoption as the default solver.}

\subsection{Comparison with Hadamard Ordering methods}

\brayan{We compare the proposed scene-driven Hadamard ordering strategy against four established ordering schemes: Sequency~\cite{agaian2011hadamard}, Zig-Zag~\cite{lopez2022efficient}, XY~\cite{cai2023detail}, and Cake-Cutting~\cite{yu2019super}. Quantitative results on EUROSAT (Table~\ref{tab:comparison2}) and ARAD (Table~\ref{tab:comparisonarad512}) show that the proposed method provides consistent improvements across datasets and sampling budgets, achieving higher PSNR/SSIM and lower SAM than fixed-order alternatives. On EUROSAT, our ordering yields the best PSNR at all reported sampling ratios ($\delta\in\{10\%,20\%,30\%\}$), reaching 36.05\,dB, 39.17\,dB, and 40.65\,dB, respectively, with gains of up to 1.98\,dB over the strongest baseline; the corresponding SSIM increases and SAM reductions indicate improved spatial quality and spectral fidelity. On the ARAD dataset, the proposed method similarly achieves state-of-the-art performance at low sampling rates, yielding PSNR gains of 1.07\,dB and 1.57\,dB at $\delta=5\%$ and $\delta=10\%$, respectively, while concurrently increasing SSIM and decreasing SAM.}

\begin{table}[!t] \centering

\caption{\textbf{Comparison results in EUROSAT}. best and second best results highlighted in {\textbf{bold}} and \underline{underline} respectively.\label{tab:comparison2}} 

\resizebox{\linewidth}{!}{%
\begin{tabular}{l|l||ccc}
\toprule
$\delta$ & Method   & PSNR ($\uparrow$)     & SSIM ($\uparrow$)    & SAM ($\downarrow$)    \\ \hline \hline 
& Sequency~\cite{agaian2011hadamard}                         & 31.28          & 0.828        & 0.0513    \\
& Zig-Zag~\cite{lopez2022efficient}  & 34.41          & 0.894        & 0.0377    \\
10\% & XY~\cite{cai2023detail}       & \ul{34.67}     & \ul{0.898}  & \ul{0.0371}   \\
& Cake Cutting~\cite{yu2019super}    & 33.88          & 0.875        &  0.0408  \\
& Ours                      & \bb{36.05}    & \bb{0.924}   & \bb{0.0332}   \\ 
\midrule 
& Sequency~\cite{agaian2011hadamard}                            & 34.16          & 0.891       &  0.0384    \\
& Zig-Zag~\cite{lopez2022efficient}  & 37.19          & 0.945       & 0.0279    \\
20\% & XY~\cite{cai2023detail}       &  37.11         & 0.943       & 0.0284 \\
& Cake Cutting~\cite{yu2019super}    & \ul{36.87}     & \ul{0.934}  & \ul{0.0302}   \\ 
& Ours                        & \bb{39.17}  & \bb{0.963}  & \bb{0.0244}   \\ 
\midrule 
& Sequency~\cite{agaian2011hadamard}                            & 36.23          & 0.932       & 0.0296    \\
& Zig-Zag~\cite{lopez2022efficient}  & 38.51          & 0.960       & 0.0237    \\
30\% & XY~\cite{cai2023detail}       & 38.69          & 0.960       & 0.0239 \\
& Cake Cutting~\cite{yu2019super}    & \ul{39.23}     & \ul{0.961} & \ul{0.0239}   \\ 
& Ours                      & \bb{40.65}    & \bb{0.973}   & \bb{0.0210}   \\ 
\bottomrule
\end{tabular}%
} \vspace{-1em}
\end{table}

\section{Real implementation performance analysis}

\begin{figure}[!t]
    \centering
    \includegraphics[width=\linewidth]{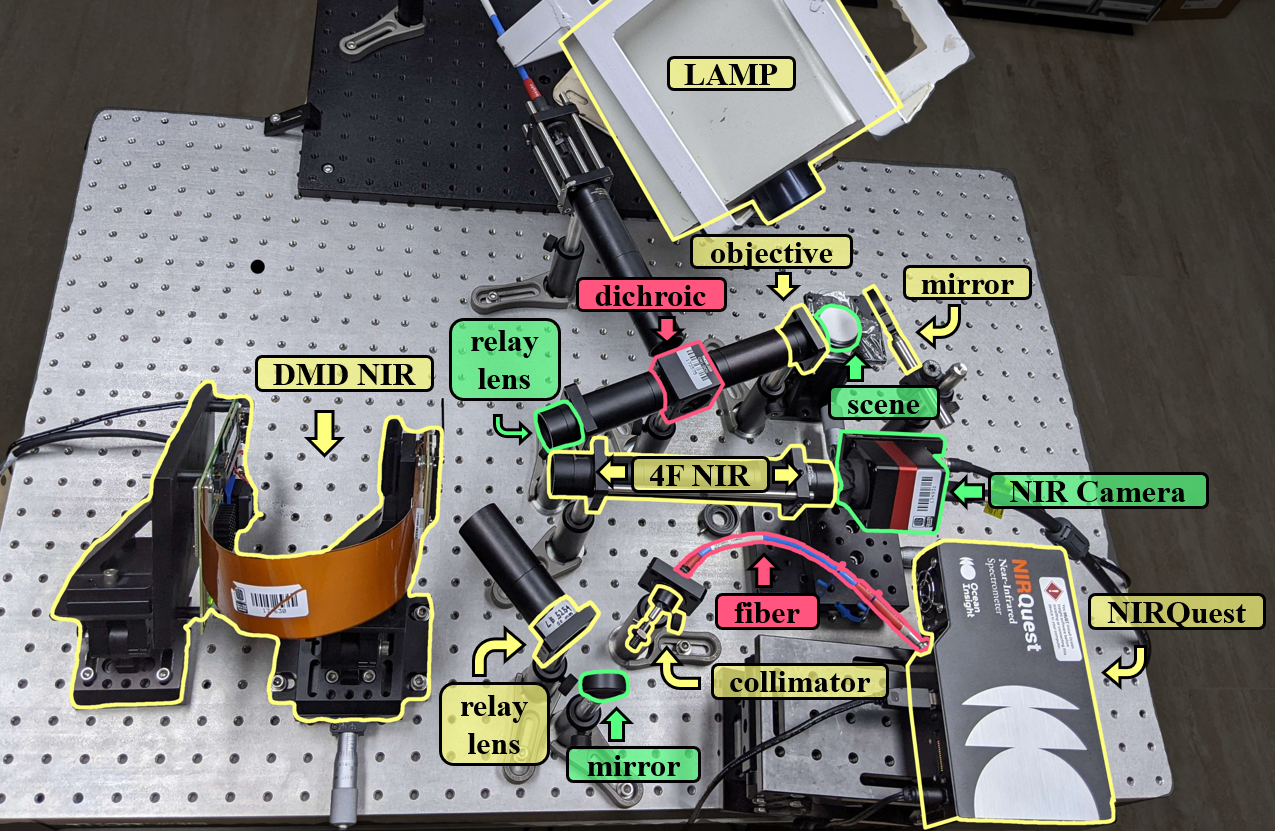}
    \caption{\textbf{Optical setup for Single Pixel Camera (SPC).} Featuring an IT 3900e infrared (IR) lamp as the light source, an IR-optimized Digital Micromirror Device (DMD) for spatial coding, and a NIRQuest spectrometer for measurement acquisition. Additional IR-optimized components such as lenses and mirrors are integrated to enhance performance.}
    \label{fig:IMPLEMENTAION} \vspace{-1.5em}
\end{figure}

To validate the effectiveness of the proposed sensing approach, it is imperative to construct an optical testbed capable of facilitating performance analysis in realistic noise scenarios. It is noteworthy that, for Near-Infrared (NIR) implementations, specialized optical elements optimized for this wavelength range are required. In this context, Figure~\ref{fig:IMPLEMENTAION} shows the testbed setup at the HDSP laboratory. It is built with a NIR lamp (3900e-Ilumination technology) that illuminates the sample scene, in which light is redirected by a broadband mirror through the objective lens and the relay lens, employing Thorlabs LB5552 bi-convex lenses, it is important to note that a dichroic mirror is used to filter the NIR spectrum, the setup then constructs an imaging plane on the DMD for the NIR range (Vialux GmbH vd65), enabling the reflection surface to encode the scene. Subsequently, another relay lens, in conjunction with a collimator lens, directs the ray lights into an optical fiber (QP1000-025-VIS-NIR). Finally, the setup includes a NIRQUEST spectrometer from Ocean Insight, which allows the acquisition of up to $512$ spectral bands. Employing the NIR SPC system, a series of 8 scenes were captured using a complete Hadamard set of patterns, serving as the ground truth for subsequent performance analysis. \brayan{For online operation, the lightweight selector model (comprising 5.5 million parameters, $\sim$20.8~MB) requires an inference time of only 1.8~ms on a GPU (14.2~ms on a CPU) to determine the scene-driven ordering, introducing a negligible computational footprint to the acquisition pipeline.}

\begin{table}[!t] \centering

\caption{\textbf{Comparison results in ARAD}. best and second best results highlighted in {\textbf{bold}} and \underline{underline} respectively. \label{tab:comparisonarad512}} 

\resizebox{\linewidth}{!}{%
\begin{tabular}{l|l||ccc}
\toprule
$\delta$ & Method   & PSNR ($\uparrow$)     & SSIM ($\uparrow$)    & SAM ($\downarrow$)    \\ \hline \hline 
& Sequency~\cite{agaian2011hadamard}  & 27.06  & 0.746 & 0.0591  \\
& Zig-Zag~\cite{lopez2022efficient}   & 32.18  & \ul{0.831} & \ul{0.0322}  \\
5\% & XY~\cite{cai2023detail}        & \ul{32.19}  & 0.830  & 0.0331  \\
& Cake Cutting~\cite{yu2019super}     & 31.39 & 0.805  & 0.0410   \\
& Ours                                & \textbf{33.26} & \textbf{0.854}  & \textbf{0.0311}   \\
\midrule 
& Sequency~\cite{agaian2011hadamard}  & 29.73  & 0.793  & 0.0462   \\
& Zig-Zag~\cite{lopez2022efficient}   & 34.27  & 0.884  & 0.0292   \\
10\% & XY~\cite{cai2023detail}        & \ul{34.69}  & \ul{0.890}  & \ul{0.0279}   \\
& Cake Cutting~\cite{yu2019super}     &  33.89  & 0.866  & 0.0321   \\
& Ours                                & \bb{36.26}  & \bb{0.913}  & \bb{0.0248}   \\
\midrule 
& Sequency~\cite{agaian2011hadamard}  & 33.21  & 0.868   & 0.0320   \\
& Zig-Zag~\cite{lopez2022efficient}   & \ul{37.27}  & \ul{0.941}   & \ul{0.0207}   \\
20\% & XY~\cite{cai2023detail}        & 37.26  & 0.940   & 0.0212  \\
& Cake Cutting~\cite{yu2019super}     & 37.26  & 0.930   & 0.0231    \\
& Ours                                & \bb{39.87}  & \bb{0.958}  & \bb{0.0193}   \\
\bottomrule
\end{tabular}%
} \vspace{-1em}
\end{table}

\begin{figure*}[!t]
    \centering
    \includegraphics[width=\linewidth]{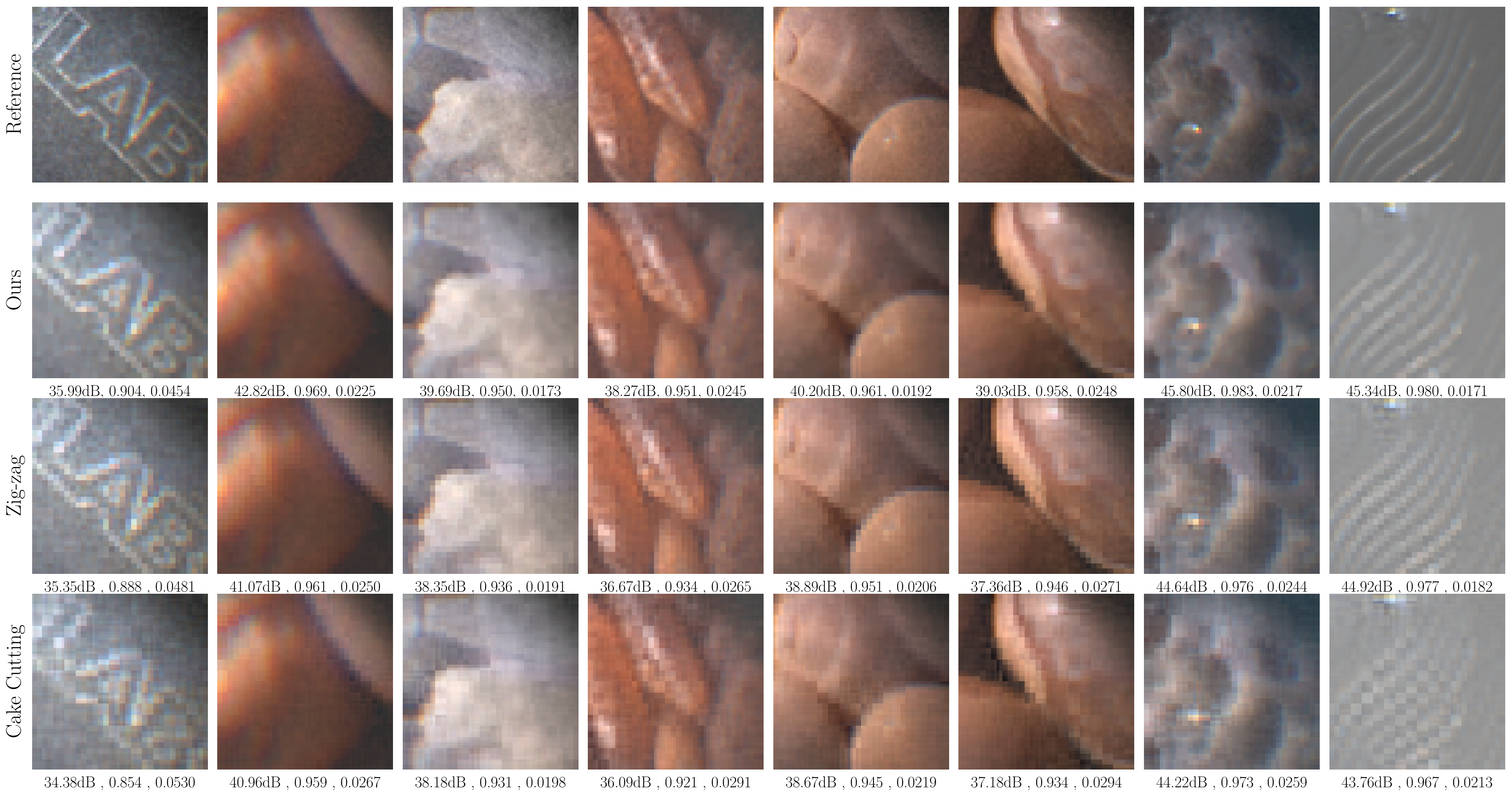} \vspace{-2em}
    \caption{Real reconstruction results of proposed method and fixed ordering methodologies on NIR spectral images. The first rows correspond to reference NIR spectral images reconstructed from a full-sensing of the Hadamard Single Pixel, and the remaining rows correspond to the proposed scene-driven sensing and Zig-zag and Cake Cutting ordering methods, respectively. \vspace{-2em}} 
    \label{fig:realresults}
\end{figure*}

\subsection{Implemented optical system acquisition protocol}

In our acquisition protocol, we sequentially obtain a set of SPC measurements 
$\{ \mathbf{y}_i \}_{i=1}^n$ corresponding to a series of modulation patterns 
$\{ \mathbf{h}_i \}_{i=1}^n$. In practical implementations, each modulation pattern in the Hadamard basis contains coefficients in $\{-1,1\}$, whereas the DMD accepts only binary values $\{0,1\}$. Using the standard complementary strategy for signed patterns on binary optical devices and balanced binary (S-type) matrices, SPC measurements decompose each Hadamard row $\mathbf{h}_i$ into positive and negative binary components, then subtract their detector responses. Accordingly, the model for each Near-Infrared (NIR) spectral response associated with a modulation pattern becomes
\begin{equation}
\mathbf{y}_i^\top = (\mathbf{h}_i^{+})^\top \mathbf{X} - (\mathbf{h}_i^{-})^\top \mathbf{X},
\end{equation}
where $\mathbf{h}_i^{+}$ and $\mathbf{h}_i^{-}$ denote the binary patterns derived from the positive and negative entries of $\mathbf{h}_i$, respectively~\cite{vaz2020image}. This complementary acquisition cancels the DC term introduced by the $\{-1,1\}\!\to\!\{0,1\}$ mapping and ensures that the effective measurement is equivalent to applying the original signed Hadamard coefficient. \brayan{In our sensing protocol, the indices $i$ are  those indicated by the support of the predicted mask $\mathbf{m}_a$. As each SPC measurement is linked to a Hadamard row $\mathbf{h}_i$, real-time reconstruction becomes straightforward through the cumulative addition of the current set of acquisitions.} Then, the cumulative estimation of the NIR spectral image in the $k$-th snapshot is expressed mathematically as
\begin{equation}
\hat{\mathbf{X}}_k = \frac{1}{n} \sum_i^k \mathbf{h}_i \otimes \mathbf{y}_i,
\end{equation}
where $\otimes$ denotes the outer product. Because optical light modulation occurs spatially, it is inherently broadcast across spectral dimensions; thus, the spectrometer’s specifications determine how many bands can be used.

The NIR spectral images, comprising eight scenes featuring diverse materials such as metal, various cereals and grains, rocks, and a coin, are captured for analysis (Figure~\ref{fig:realresults}). Hadamard coefficients are collected for each scene to facilitate the emulation of different Hadamard orderings and the proposed methodology. In Figure~\ref{fig:realresults}, we compare several ordering methods with the proposed approach at a $25\%$ sampling rate. The Hadamard coefficient estimation model uses parameters trained on EuroSAT at the same sampling rate, requiring no additional training. The proposed method outperforms all others for every scene, with Zig-Zag and Cake Cutting alternately ranking second, consistent with the simulation trends. Trained on remote sensing spectral images, the model generalizes well to spatial information, delivering acceptable image quality for Optical Laboratory images. \vspace{-0.4em}

\section{Conclusions}

\brayan{This work presents a deep scene-driven sensing protocol designed to efficiently capture spectral images with high reconstruction quality while minimizing the number of modulation patterns required in HSPI. By combining an initial predetermined acquisition stage with a learned, binary classification-based selection mechanism that predicts the most informative remaining Hadamard coefficients, the proposed approach preserves DMD-friendly Hadamard $\{\pm1\}$ modulation while enabling scene-driven adaptive ordering. Experimental results on simulated datasets and a real single-pixel spectral imaging testbed validate the effectiveness of the method, consistently outperforming conventional fixed Hadamard orderings across sampling regimes in terms of PSNR/SSIM and SAM. Overall, the proposed sequential sensing and recovery strategy provides a practical pathway to reduce acquisition time and improve spectral fidelity for visible and infrared single-pixel spectral imaging under constrained snapshot budgets.}

\section{Acknowledgments}

The authors extend their sincere appreciation to Ana Karina for her invaluable support in the testbed implementation, which greatly contributed to the success of this research. Furthermore, we acknowledge the VIE-UIS under project 3924, which supports the realization of this study. \vspace{-1em}

\vspace{11pt}

{\small
\bibliographystyle{IEEEtran}
\bibliography{references}

\begin{thebibliography}{10}
\providecommand{\url}[1]{#1}
\csname url@samestyle\endcsname
\providecommand{\newblock}{\relax}
\providecommand{\bibinfo}[2]{#2}
\providecommand{\BIBentrySTDinterwordspacing}{\spaceskip=0pt\relax}
\providecommand{\BIBentryALTinterwordstretchfactor}{4}
\providecommand{\BIBentryALTinterwordspacing}{\spaceskip=\fontdimen2\font plus
\BIBentryALTinterwordstretchfactor\fontdimen3\font minus
  \fontdimen4\font\relax}
\providecommand{\BIBforeignlanguage}[2]{{%
\expandafter\ifx\csname l@#1\endcsname\relax
\typeout{** WARNING: IEEEtran.bst: No hyphenation pattern has been}%
\typeout{** loaded for the language `#1'. Using the pattern for}%
\typeout{** the default language instead.}%
\else
\language=\csname l@#1\endcsname
\fi
#2}}
\providecommand{\BIBdecl}{\relax}
\BIBdecl

\bibitem{garini2006spectral}
Y.~Garini, I.~T. Young, and G.~McNamara, ``Spectral imaging: principles and
  applications,'' \emph{Cytometry Part A: The Journal of the International
  Society for Analytical Cytology}, vol.~69, no.~8, pp. 735--747, 2006.

\bibitem{bacca2023computational}
J.~Bacca, E.~Martinez, and H.~Arguello, ``Computational spectral imaging: a
  contemporary overview,'' \emph{JOSA A}, vol.~40, no.~4, pp. C115--C125, 2023.

\bibitem{shaw2003spectral}
G.~A. Shaw and H.~K. Burke, ``Spectral imaging for remote sensing,''
  \emph{Lincoln laboratory journal}, vol.~14, no.~1, pp. 3--28, 2003.

\bibitem{teena2014thermal}
M.~Teena and A.~Manickavasagan, ``Thermal infrared imaging,'' in \emph{Imaging
  with Electromagnetic Spectrum: Applications in Food and Agriculture}.\hskip
  1em plus 0.5em minus 0.4em\relax Springer, 2014, pp. 147--173.

\bibitem{zhu2018review}
L.~Zhu, J.~Suomalainen, J.~Liu, J.~Hyypp{\"a}, H.~Kaartinen, H.~Haggren
  \emph{et~al.}, ``A review: Remote sensing sensors,'' \emph{Multi-purposeful
  application of geospatial data}, pp. 19--42, 2018.

\bibitem{rogalski2022scaling}
A.~Rogalski, ``Scaling infrared detectors—status and outlook,'' \emph{Reports
  on Progress in Physics}, vol.~85, no.~12, p. 126501, 2022.

\bibitem{rogalski2016challenges}
A.~Rogalski, P.~Martyniuk, and M.~Kopytko, ``Challenges of small-pixel infrared
  detectors: a review,'' \emph{Reports on Progress in Physics}, vol.~79, no.~4,
  p. 046501, 2016.

\bibitem{duarte2008single}
M.~F. Duarte, M.~A. Davenport, D.~Takhar, J.~N. Laska, T.~Sun, K.~F. Kelly, and
  R.~G. Baraniuk, ``Single-pixel imaging via compressive sampling,'' \emph{IEEE
  signal processing magazine}, vol.~25, no.~2, pp. 83--91, 2008.

\bibitem{sampsell1994digital}
J.~B. Sampsell, ``Digital micromirror device and its application to projection
  displays,'' \emph{Journal of Vacuum Science \& Technology B: Microelectronics
  and Nanometer Structures Processing, Measurement, and Phenomena}, vol.~12,
  no.~6, pp. 3242--3246, 1994.

\bibitem{centrone2015infrared}
A.~Centrone, ``Infrared imaging and spectroscopy beyond the diffraction
  limit,'' \emph{Annual review of analytical chemistry}, vol.~8, pp. 101--126,
  2015.

\bibitem{gibson2020single}
G.~M. Gibson, S.~D. Johnson, and M.~J. Padgett, ``Single-pixel imaging 12 years
  on: a review,'' \emph{Optics express}, vol.~28, no.~19, pp. 28\,190--28\,208,
  2020.

\bibitem{garcia2020optimized}
H.~Garcia, C.~V. Correa, and H.~Arguello, ``Optimized sensing matrix for single
  pixel multi-resolution compressive spectral imaging,'' \emph{IEEE
  Transactions on Image Processing}, vol.~29, pp. 4243--4253, 2020.

\bibitem{monroy2023deep}
B.~Monroy, J.~Bacca, and H.~Arguello, ``Deep adaptive superpixels for hadamard
  single pixel imaging in near-infrared spectrum,'' in \emph{ICASSP 2023-2023
  IEEE International Conference on Acoustics, Speech and Signal Processing
  (ICASSP)}.\hskip 1em plus 0.5em minus 0.4em\relax IEEE, 2023, pp. 1--5.

\bibitem{bacca2020coupled}
J.~Bacca, L.~Galvis, and H.~Arguello, ``Coupled deep learning coded aperture
  design for compressive image classification,'' \emph{Optics express},
  vol.~28, no.~6, pp. 8528--8540, 2020.

\bibitem{vaz2020image}
P.~G. Vaz, D.~Amaral, L.~R. Ferreira, M.~Morgado, and J.~Cardoso, ``Image
  quality of compressive single-pixel imaging using different hadamard
  orderings,'' \emph{Optics express}, vol.~28, no.~8, pp. 11\,666--11\,681,
  2020.

\bibitem{zhang2017hadamard}
Z.~Zhang, X.~Wang, G.~Zheng, and J.~Zhong, ``Hadamard single-pixel imaging
  versus fourier single-pixel imaging,'' \emph{Optics Express}, vol.~25,
  no.~16, pp. 19\,619--19\,639, 2017.

\bibitem{agaian2011hadamard}
S.~S. Agaian, H.~Sarukhanyan, K.~Egiazarian, and J.~Astola, ``Hadamard
  transforms.''\hskip 1em plus 0.5em minus 0.4em\relax SPIE, 2011.

\bibitem{yu2019super}
W.-K. Yu, ``Super sub-nyquist single-pixel imaging by means of cake-cutting
  hadamard basis sort,'' \emph{Sensors}, vol.~19, no.~19, p. 4122, 2019.

\bibitem{sun2017russian}
M.-J. Sun, L.-T. Meng, M.~P. Edgar, M.~J. Padgett, and N.~Radwell, ``A russian
  dolls ordering of the hadamard basis for compressive single-pixel imaging,''
  \emph{Scientific reports}, vol.~7, no.~1, p. 3464, 2017.

\bibitem{yu2020super}
X.~Yu, R.~I. Stantchev, F.~Yang, and E.~Pickwell-MacPherson, ``Super
  sub-nyquist single-pixel imaging by total variation ascending ordering of the
  hadamard basis,'' \emph{Scientific Reports}, vol.~10, no.~1, p. 9338, 2020.

\bibitem{lopez2022efficient}
L.~L{\'o}pez-Garc{\'\i}a, W.~Cruz-Santos, A.~Garc{\'\i}a-Arellano,
  P.~Filio-Aguilar, J.~A. Cisneros-Mart{\'\i}nez, and R.~Ramos-Garc{\'\i}a,
  ``Efficient ordering of the hadamard basis for single pixel imaging,''
  \emph{Optics Express}, vol.~30, no.~8, pp. 13\,714--13\,732, 2022.

\bibitem{cai2022detail}
Y.~Cai, S.~Li, W.~Zhang, H.~Wu, X.-r. Yao, and Q.~Zhao, ``A detail-enhanced
  sampling strategy in hadamard single-pixel imaging,'' \emph{arXiv preprint
  arXiv:2209.04449}, 2022.

\bibitem{huang2022spectral}
L.~Huang, R.~Luo, X.~Liu, and X.~Hao, ``Spectral imaging with deep learning,''
  \emph{Light: Science \& Applications}, vol.~11, no.~1, p.~61, 2022.

\bibitem{choi2017}
\BIBentryALTinterwordspacing
I.~Choi, D.~S. Jeon, G.~Nam, D.~Gutierrez, and M.~H. Kim, ``High-quality
  hyperspectral reconstruction using a spectral prior,'' vol.~36, no.~6, nov
  2017. [Online]. Available: \url{https://doi.org/10.1145/3130800.3130810}
\BIBentrySTDinterwordspacing

\bibitem{zheng2021deep}
S.~Zheng, Y.~Liu, Z.~Meng, M.~Qiao, Z.~Tong, X.~Yang, S.~Han, and X.~Yuan,
  ``Deep plug-and-play priors for spectral snapshot compressive imaging,''
  \emph{Photonics Research}, vol.~9, no.~2, pp. B18--B29, 2021.

\bibitem{wang2019hyperspectral}
L.~Wang, C.~Sun, Y.~Fu, M.~H. Kim, and H.~Huang, ``Hyperspectral image
  reconstruction using a deep spatial-spectral prior,'' in \emph{Proceedings of
  the IEEE/CVF Conference on Computer Vision and Pattern Recognition}, 2019,
  pp. 8032--8041.

\bibitem{arguello2023deep}
H.~Arguello, J.~Bacca, H.~Kariyawasam, E.~Vargas, M.~Marquez, R.~Hettiarachchi,
  H.~Garcia, K.~Herath, U.~Haputhanthri, B.~S. Ahluwalia \emph{et~al.}, ``Deep
  optical coding design in computational imaging: a data-driven framework,''
  \emph{IEEE Signal Processing Magazine}, vol.~40, no.~2, pp. 75--88, 2023.

\bibitem{bacca2021deep}
J.~Bacca, T.~Gelvez-Barrera, and H.~Arguello, ``Deep coded aperture design: An
  end-to-end approach for computational imaging tasks,'' \emph{IEEE
  Transactions on Computational Imaging}, vol.~7, pp. 1148--1160, 2021.

\bibitem{hinojosa2022privhar}
C.~Hinojosa, M.~Marquez, H.~Arguello, E.~Adeli, L.~Fei-Fei, and J.~C. Niebles,
  ``Privhar: Recognizing human actions from privacy-preserving lens,'' in
  \emph{European Conference on Computer Vision}.\hskip 1em plus 0.5em minus
  0.4em\relax Springer, 2022, pp. 314--332.

\bibitem{arguello2022optics}
P.~Arguello, J.~Lopez, C.~Hinojosa, and H.~Arguello, ``Optics lens design for
  privacy-preserving scene captioning,'' in \emph{2022 IEEE International
  Conference on Image Processing (ICIP)}.\hskip 1em plus 0.5em minus
  0.4em\relax IEEE, 2022, pp. 3551--3555.

\bibitem{cai2023detail}
Y.~Cai, S.~Li, W.~Zhang, H.~Wu, X.~Yao, and Q.~Zhao, ``A detail-enhanced
  sampling strategy in hadamard single-pixel imaging,'' \emph{Chinese Optics
  Letters}, vol.~21, no.~7, p. 071101, 2023.

\bibitem{arguello2021shift}
H.~Arguello, S.~Pinilla, Y.~Peng, H.~Ikoma, J.~Bacca, and G.~Wetzstein,
  ``Shift-variant color-coded diffractive spectral imaging system,''
  \emph{Optica}, vol.~8, no.~11, pp. 1424--1434, 2021.

\bibitem{chakrabarti2016learning}
A.~Chakrabarti, ``Learning sensor multiplexing design through
  back-propagation,'' \emph{Advances in Neural Information Processing Systems},
  vol.~29, 2016.

\bibitem{diaz2018adaptive}
N.~Diaz, H.~Rueda, and H.~Arguello, ``Adaptive filter design via a gradient
  thresholding algorithm for compressive spectral imaging,'' \emph{Applied
  Optics}, vol.~57, no.~17, pp. 4890--4900, 2018.

\bibitem{dekel2008adaptive}
S.~Dekel, ``Adaptive compressed image sensing based on wavelet-trees,''
  \emph{preprint}, 2008.

\bibitem{averbuch2012adaptive}
A.~Averbuch, S.~Dekel, and S.~Deutsch, ``Adaptive compressed image sensing
  using dictionaries,'' \emph{SIAM Journal on Imaging Sciences}, vol.~5, no.~1,
  pp. 57--89, 2012.

\bibitem{radwell2014single}
N.~Radwell, K.~J. Mitchell, G.~M. Gibson, M.~P. Edgar, R.~Bowman, and M.~J.
  Padgett, ``Single-pixel infrared and visible microscope,'' \emph{Optica},
  vol.~1, no.~5, pp. 285--289, 2014.

\bibitem{monroy2024predicting}
B.~Monroy, J.~Bacca, and H.~Arguello, ``Predicting the spectrum: Deep adaptive
  sensing for hadamard single pixel spectral imaging,'' in \emph{2024 IEEE
  International Conference on Acoustics, Speech, and Signal Processing
  Workshops (ICASSPW)}.\hskip 1em plus 0.5em minus 0.4em\relax IEEE, 2024, pp.
  154--158.

\bibitem{mallat1999wavelet}
S.~Mallat, \emph{A wavelet tour of signal processing}.\hskip 1em plus 0.5em
  minus 0.4em\relax Elsevier, 1999.

\bibitem{chan2016plug}
S.~H. Chan, X.~Wang, and O.~A. Elgendy, ``Plug-and-play admm for image
  restoration: Fixed-point convergence and applications,'' \emph{IEEE
  Transactions on Computational Imaging}, vol.~3, no.~1, pp. 84--98, 2016.

\bibitem{aggarwal2016hyperspectral}
H.~K. Aggarwal and A.~Majumdar, ``Hyperspectral image denoising using
  spatio-spectral total variation,'' \emph{IEEE Geoscience and Remote Sensing
  Letters}, vol.~13, no.~3, pp. 442--446, 2016.

\bibitem{buzzard2018plug}
G.~T. Buzzard, S.~H. Chan, S.~Sreehari, and C.~A. Bouman, ``Plug-and-play
  unplugged: Optimization-free reconstruction using consensus equilibrium,''
  \emph{SIAM Journal on Imaging Sciences}, vol.~11, no.~3, pp. 2001--2020,
  2018.

\bibitem{venkatakrishnan2013plug}
S.~V. Venkatakrishnan, C.~A. Bouman, and B.~Wohlberg, ``Plug-and-play priors
  for model based reconstruction,'' in \emph{2013 IEEE global conference on
  signal and information processing}.\hskip 1em plus 0.5em minus 0.4em\relax
  IEEE, 2013, pp. 945--948.

\bibitem{elad2023image}
M.~Elad, B.~Kawar, and G.~Vaksman, ``Image denoising: The deep learning
  revolution and beyond—a survey paper,'' \emph{SIAM Journal on Imaging
  Sciences}, vol.~16, no.~3, pp. 1594--1654, 2023.

\bibitem{tachella2025deepinverse}
J.~Tachella, M.~Terris, S.~Hurault, A.~Wang, D.~Chen, M.-H. Nguyen, M.~Song,
  T.~Davies, L.~Davy, J.~Dong \emph{et~al.}, ``Deepinverse: A python package
  for solving imaging inverse problems with deep learning,'' \emph{arXiv
  preprint arXiv:2505.20160}, 2025.

\bibitem{wang2021single}
F.~Wang, C.~Wang, C.~Deng, S.~Han, and G.~Situ, ``Single-pixel imaging using
  physics enhanced deep learning,'' \emph{Photonics Research}, vol.~10, no.~1,
  pp. 104--110, 2021.

\bibitem{higham2018deep}
C.~F. Higham, R.~Murray-Smith, M.~J. Padgett, and M.~P. Edgar, ``Deep learning
  for real-time single-pixel video,'' \emph{Scientific reports}, vol.~8, no.~1,
  p. 2369, 2018.

\bibitem{hahamovich2021single}
E.~Hahamovich, S.~Monin, Y.~Hazan, and A.~Rosenthal, ``Single pixel imaging at
  megahertz switching rates via cyclic hadamard masks,'' \emph{Nature
  communications}, vol.~12, no.~1, p. 4516, 2021.

\end{thebibliography}
}

\clearpage
\onecolumn

\begin{center}
    \section*{Supplemental Materials}
\end{center}

\section{Additional Simulations.}

\subsection{Evaluation of loss functions}

In this section, we comprehensively evaluate the proposed method under different training strategies. Specifically, we compare the proposed \textbf{Sparse} loss against two alternatives, \textbf{Data Fidelity} loss and \textbf{Ordering} loss. We describe each loss below.

\textbf{Sparse Loss.} The proposed sparse loss follows the binary classification formulation used in the main manuscript. First, an oracle sparse support mask $\mathbf{m}_s \in \{0,1\}^{n}$ is constructed from the full Hadamard spectrum by selecting the top-$k_a$ non-predetermined coefficients according to their aggregated magnitude $|\mathbf{Y}|_{\Sigma}$. Then, the selector network $\mathcal{P}_{\theta}$ receives the predetermined measurements $\mathbf{Y}_p$ and predicts a score map over the Hadamard spectrum. The sparse loss, denoted as $\mathcal{L}_s$, is defined as the binary cross-entropy between the oracle support $\mathbf{m}_s$ and the predicted mask:
\begin{equation}
\mathcal{L}_s(\mathbf{m}_s,\mathbf{Y}_p)
=
\mathrm{BCE}
\left[
\mathbf{m}_s,
\mathcal{P}_{\theta}(\mathbf{Y}_p)
\right].
\end{equation}
This loss directly enforces the selection of the most informative Hadamard coefficients through the sparse oracle support, whose cardinality is fixed by the desired scene-driven sampling budget $k_a$. Therefore, the compression ratio is controlled by construction, without requiring an additional transmittance regularization term.

\textbf{Data Fidelity Loss.} In the ``data-fidelity'' loss, the primary objective is to minimize the reconstruction error between the estimated image and the ground-truth image while encouraging the desired amount of modulation patterns. Let $\delta_a=k_a/n$ denote the target scene-driven sampling ratio. Following the DOCD framework, this strategy uses a two-term cost function composed of a reconstruction error and a transmittance regularization on the sensing mask $\mathbf{m}_a$. This regularization encourages the average mask transmittance to match the desired compression ratio. In this sense, the loss function for this perspective, denoted as $\mathcal{L}_f$, is given by
\begin{equation}
\mathcal{L}_f(\mathbf{X}, \hat{\mathbf{X}}, \mathbf{m}_a)
=
\Vert \mathbf{X} - \hat{\mathbf{X}} \Vert_F^2
+
\left\Vert
\frac{1}{n}\mathbf{1}^{\top}\mathbf{m}_a - \delta_a
\right\Vert_2^2.
\end{equation}

\textbf{Ordering Loss.} The ``ordering'' loss consists of an index regression training directly between the estimated measurement mask $\mathbf{m}_a$ and the indexed sparse ordering vector $\mathbf{m}_o = \texttt{argsort}(|\mathbf{Y}|_{\Sigma})$. This approach does not explicitly constrain a specific transmittance level, as it learns the entry ordering of the coefficients for the spectral images. The loss function for this perspective, denoted as $\mathcal{L}_o$, is based on the mean squared error:
\begin{equation}
\mathcal{L}_o(\mathbf{m}_a, \mathbf{m}_o)
=
\Vert \mathbf{m}_a - \mathbf{m}_o \Vert_2^2.
\end{equation}

\begin{figure*}[!h]
    \centering
    \includegraphics[width=\linewidth]{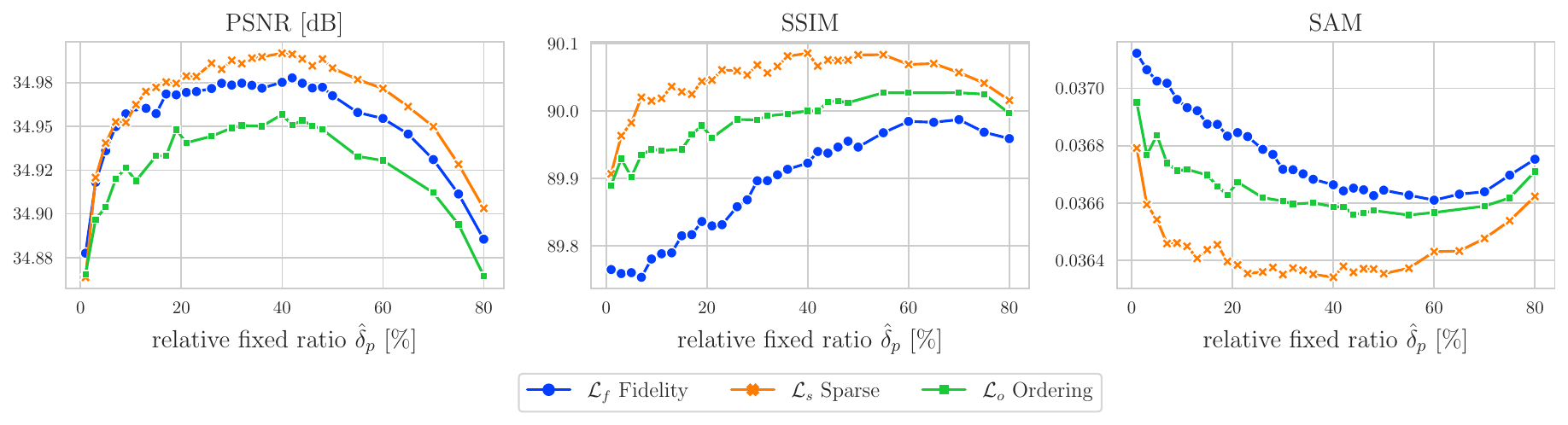}
    \caption{Performance of different training cost functions for different split sensing configurations, relative predetermined ratio is defined as $\hat{\delta}_p = \delta_p / m $.} 
    \label{fig:training}
\end{figure*}

Analysis and discussion are based on Figure~\ref{fig:training}. The performance of the proposed method is evaluated across several metrics. The \textit{fidelity} loss secures the second-best PSNR but performs lower in SSIM and SAM, especially for low $\delta_p$. In contrast, the \textit{sparse} loss performs better in all scenarios and metrics. The ordering loss places third in the PSNR ranking. Notably, the \textbf{fidelity} loss requires two optimization terms, which are difficult to tune, while both \textbf{sparse} and \textbf{ordering} use a single optimization term. However, the sparse BCE loss directly satisfies the desired compression ratio through the fixed-cardinality oracle support and provides better fidelity consistency in the Hadamard spectrum. These findings highlight each strategy's strengths and limitations, with the \textit{sparse} loss being the most effective and therefore selected for the following simulations

\subsection{Influence of relative predetermined ratio $\hat{\delta}_p$}

\begin{figure*}[!h]
    \centering
    \includegraphics[width=\linewidth]{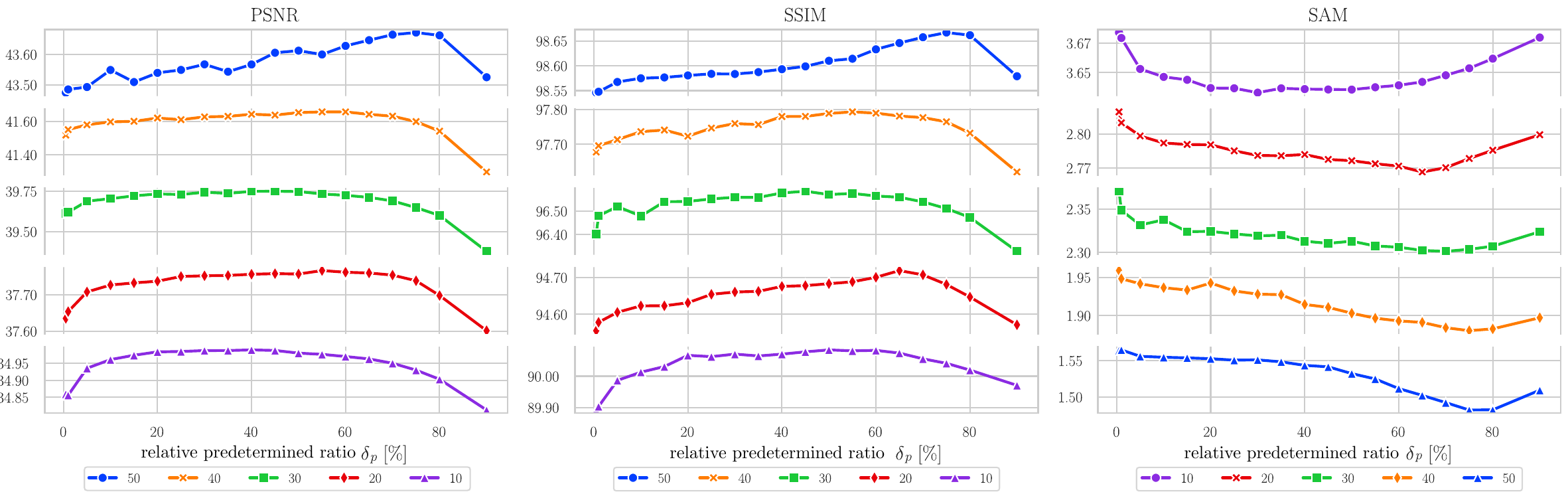}
    \caption{Relative performance on PSNR, SSIM, and SAM scores under different total sampling ratios $\delta$ (color bars) and relative predetermined ratios $\hat{\delta}_p$ (x-axis). The relative predetermined ratio is defined as $\hat{\delta}_p = k_p/k = \delta_p/\delta$.}
    \label{fig:perform}
\end{figure*}

We explore the influence of the relative predetermined ratio
$\hat{\delta}_p = k_p/k = \delta_p/\delta$, where $k_p$ is the number of predetermined coefficients, $k=k_p+k_a$ is the total number of sensed coefficients, $\delta_p=k_p/n$ is the predetermined sampling ratio, and $\delta=k/n$ is the total sampling ratio. We consider total sampling ratios $\delta$ ranging from $10\%$ to $50\%$. For each compression configuration, $\hat{\delta}_p$ is evaluated from $1\%$ to $95\%$, and the number of scene-driven coefficients is given by $k_a=k-k_p$. Spatial and spectral metrics are shown in Figure~\ref{fig:perform}.

Our experiments show a trade-off between predetermined and scene-driven sensing. The optimal value of $\hat{\delta}_p$ is typically between $50\%$ and $80\%$ as the total sampling ratio increases. A critical threshold occurs when $\hat{\delta}_p$ exceeds $80\%$, where most sensed coefficients come from the fixed predetermined ordering and less than $20\%$ of the budget remains for scene-driven measurements. This balance is crucial for optimal performance. Conversely, performance improves significantly when $\hat{\delta}_p$ increases from very small values to the range of $1\%$--$20\%$, suggesting that a minimum number of predetermined coefficients is necessary to capture coarse scene information and guide the subsequent scene-driven selection.

We report accuracy and F1-score between the reference sparse coefficients (orange) and the scene-driven selected coefficients (yellow) as mask-level metrics. Simulations show approximately $91\%$ accuracy and $70\%$ F1-score. The high accuracy may be partially explained by class imbalance, since the total sampling ratios are below $50\%$. An F1-score of $70\%$ indicates satisfactory support recovery, while also leaving room for further refinement of the scene-driven Hadamard coefficient selection. These results suggest that coefficient-level comparison is a useful benchmark for future scene-driven Hadamard sensing research. Depending on each dataset's spatial distribution, analyzing the F1-score in the Hadamard spectrum can help establish thresholds for when scene-driven sensing is preferable to fixed ordering strategies.

\subsection{Additional Reconstruction Results}
Table~\ref{tab:comparison2} reports additional EUROSAT reconstruction results for three sampling rates ($\delta\in\{10\%,20\%,30\%\}$) under two reconstruction strategies: the transpose (backprojection) and the consensus equilibrium solver. Across all sampling rates, the proposed data-driven ordering consistently outperforms the fixed Hadamard orderings in PSNR and SSIM, while also reducing SAM, indicating improved spectral fidelity. At $\delta=10\%$, our method achieves the best performance for both transpose and consensus, providing gains over the strongest baseline (XY) on all three metrics. At $\delta=20\%$, our method attains the best transpose results and remains competitive under consensus, matching the top SSIM and achieving SAM within the top two methods. At $\delta=30\%$, our method again yields the best results in both reconstruction settings, with the largest improvements observed in the transpose case, suggesting that the learned ordering particularly benefits low-complexity reconstruction. Overall, the results validate that adaptive selection of informative Hadamard coefficients improves reconstruction quality.

\begin{table}[H] \centering
\caption{\textbf{Comparison results in EUROSAT}. Best and second best results highlighted in {\textbf{bold}} and \underline{underline} respectively.\label{tab:comparison2}} 

\resizebox{0.9\linewidth}{!}{%
\begin{tabular}{l|l||ccc|ccc}
\toprule
$\delta$ & Method   
& \multicolumn{3}{c|}{Transpose} 
& \multicolumn{3}{c}{Consensus} \\
\cline{3-8}
& 
& PSNR ($\uparrow$) & SSIM ($\uparrow$) & SAM ($\downarrow$) 
& PSNR ($\uparrow$) & SSIM ($\uparrow$) & SAM ($\downarrow$) \\ 
\hline \hline 
& Sequency          & 31.28          & 0.828        & 0.0513    
                    & 31.47          & 0.836        & 0.0505         \\
& Zig-Zag           & 34.41          & 0.894        & 0.0377    
                    & 35.66          & 0.922        & 0.0339         \\
10\% & XY            & \ul{34.67}    & \ul{0.898}   & \ul{0.0371}   
                    & \ul{35.89}     & \ul{0.924}   & \ul{0.0333}         \\
& Cake Cutting      & 33.88          & 0.875        & 0.0408  
                    & 34.64          & 0.898        & 0.0380         \\
& Ours              & \bb{34.98}     & \bb{0.900}   & \bb{0.0363}        
                    & \bb{36.05}     & \bb{0.924}   & \bb{0.0332}  \\
\midrule 
& Sequency          & 34.16          & 0.891        & 0.0384    
                    & 34.64          & 0.903        & 0.0370        \\
& Zig-Zag           & 37.19          & 0.945        & 0.0279    
                    & 39.08          & 0.962        & 0.0247         \\
20\% & XY           & 37.11          & 0.941        & 0.0284 
                    & \ul{39.13}     & \ul{0.963}   & \bb{0.0242}        \\
& Cake Cutting      & \ul{36.87}     & \ul{0.934}   & \ul{0.0302}   
                    & 38.02          & 0.951        & 0.0274         \\ 
& Ours              & \bb{37.77}     & \bb{0.948}   & \bb{0.0276}         
                    & \bb{39.17}     & \bb{0.963}   & \ul{0.0244}  \\ 
\midrule 
& Sequency          & 36.23          & 0.932        & 0.0296    
                    & 36.99          & 0.944        & 0.0281         \\
& Zig-Zag           & 38.51          & 0.960        & 0.0237    
                    & \ul{40.12}     & \ul{0.968}        & 0.0215         \\
30\% & XY           & 38.69          & 0.960        & 0.0239 
                    & 40.06          & 0.950        & 0.0216         \\
& Cake Cutting      & \ul{39.23}     & \ul{0.961}   & \ul{0.0239}   
                    & 40.43          & 0.967        & 0.0215         \\ 
& Ours               & \bb{39.76}    & \bb{0.966}    & \bb{0.0229}         
                 & \bb{40.65}     & \bb{0.973}   & \bb{0.0210}  \\ 
\bottomrule
\end{tabular}%
} 
\end{table}

\section{Comparison with Fixed Learned-Pattern Baselines}
\label{sec:learned_vs_ours}

To evaluate the proposed scene-driven framework against deep learned fixed matrices, we benchmark our method against the physics-enhanced SPI strategy by Wang \emph{et al.}~\cite{wang2021single}, which employs a static bank of optimized binary patterns. For a fair comparison, we directly utilize the original 1,024 learned binary modulation patterns provided by the authors, which were optimized at a $128 \times 128$ spatial resolution ($6.25\%$ sampling budget) for facial reconstruction on the CelebA dataset. This evaluation on the baseline's native dataset allows us to directly utilize the original authors' authorized, fully-optimized pattern bank, thereby ensuring a mathematically faithful evaluation that eliminates implementation bias or sub-optimal retraining artifacts. Accordingly, we retrained our proposed selector network ($\mathcal{P}_\theta$) from scratch using the exact same CelebA training set and budget ($k_f+k_a = 1024$). While the baseline applies a uniform pattern set to all instances, our approach dynamically adapts the acquisition path by selecting a scene-dependent subset of the orthogonal Hadamard basis conditioned on initial low-frequency measurements.

This methodological distinction heavily influences reconstruction stability and solver flexibility. Because unconstrained optimization of static binary patterns disrupts standard matrix structures, the baseline lacks a meaningful algebraic relationship with simple linear operators. As shown in Table~\ref{tab:learned_vs_ours}, under a direct Transpose (backprojection) recovery, the baseline's mathematical structure collapses entirely, yielding a degenerate reconstruction ($-20.12$~dB PSNR and $0.000$ SSIM). Conversely, our framework preserves the native orthogonality of the Hadamard basis, guaranteeing stable, interpretable, and instantaneous linear initializations ($24.91$~dB PSNR and $0.689$ SSIM). 

\begin{figure}[H]
    \centering
    \includegraphics[width=0.8\linewidth]{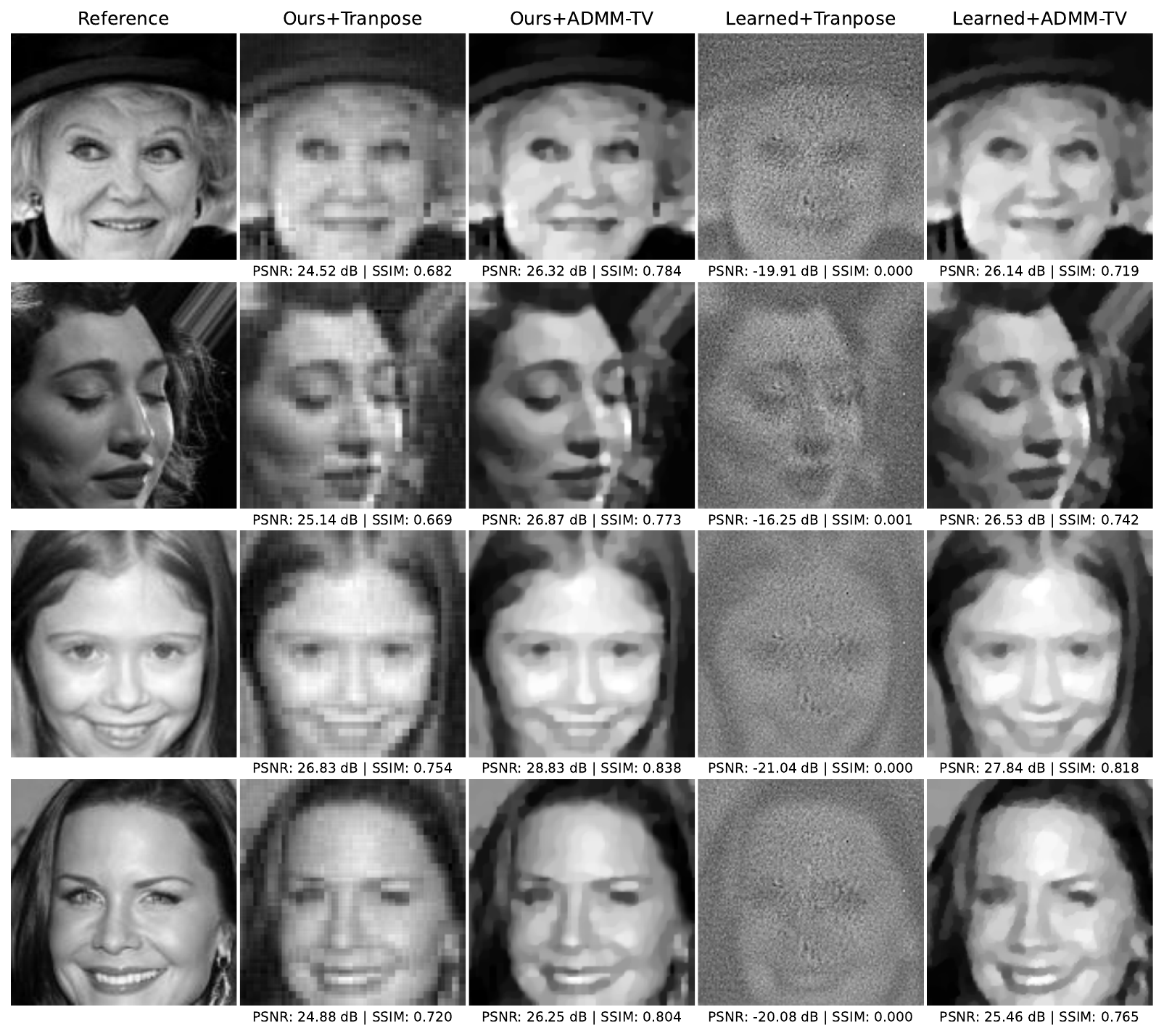}
    \caption{Visual reconstruction comparison on CelebA test samples under a $6.25\%$ acquisition budget. The static baseline~\cite{wang2021single} suffers structural collapse under direct Transpose recovery, while the proposed scene-driven Hadamard ordering ensures stable linear initializations and delivers superior fine-scale feature preservation when paired with an ADMM-TV solver.}
    \label{fig:learned_vs_ours}
\end{figure}

When paired with a sophisticated iterative solver (ADMM-TV), both methods experience an upward shift in fidelity. Nevertheless, the proposed scene-driven strategy maintains its performance edge, improving the baseline's metrics from $26.64$~dB / $0.772$ SSIM to $\mathbf{26.84}$~\textbf{dB} / $\mathbf{0.794}$ \textbf{SSIM}. Qualitative results in Figure~\ref{fig:learned_vs_ours} visually confirm these findings; the baseline exhibits severe artifacts and structural breakdown under linear inversion, whereas our approach consistently delivers superior texture preservation and sharper boundaries across both linear and iterative solvers.

\begin{table}[H]
\centering
\caption{Quantitative comparison (PSNR and SSIM) between the static learned binary baseline~\cite{wang2021single} and the proposed scene-driven Hadamard framework on CelebA at a $6.25\%$ sampling rate.}
\label{tab:learned_vs_ours}
\resizebox{0.8\linewidth}{!}{%
\begin{tabular}{llcc}
\toprule
\textbf{Sensing Strategy} & \textbf{Reconstruction Solver} & \textbf{PSNR [dB]} & \textbf{SSIM} \\
\midrule
Learned Fixed Pattern~\cite{wang2021single} & Transpose & -20.12 & 0.000 \\
Learned Fixed Pattern~\cite{wang2021single} & ADMM-TV   & 26.64  & 0.772 \\
 Proposed Scene-Driven (Ours) & Transpose & \textbf{24.91}  & \textbf{0.689} \\
 Proposed Scene-Driven (Ours) & ADMM-TV   & \textbf{26.84}  & \textbf{0.794} \\
\bottomrule
\end{tabular}
}
\end{table}

\section{Real Test-bed Implementation}

\subsection{Practical Implementation.}

In the NIR test-bed, the DMD is driven by rows of a Hadamard matrix, implemented as complementary binary patterns to realize $\{\pm 1\}$ modulation. In a practical adaptive deployment, the DMD first projects a predetermined block of low-frequency Hadamard patterns; the corresponding measurements are fed to the selector network, which outputs a binary support over the remaining Hadamard coefficients, and the second-stage modulation sequence is obtained by projecting the Hadamard patterns whose indices belong to this support. For the experiments reported here, we acquire once the complete set of Hadamard coefficients for each scene and then numerically permute them according to each ordering (fixed and data-driven), which is equivalent to executing the corresponding ordered sequences on the DMD and enables a fair comparison under identical photon and noise conditions.

\subsection{Description of NIR Test Scenes}

For completeness, we briefly describe the nine NIR scenes acquired with the SPC testbed. The scenes were designed to span a range of materials with distinct spectral signatures, including metals, organic matter, and minerals, in order to assess the robustness and generalization capability of the proposed sensing strategy.

\begin{table}[!t]
\centering
\caption{Description of NIR test scenes.}
\begin{tabular}{c l p{0.5\linewidth}}
\toprule
Scene & Material & Description \\ \midrule
1 & Matte laminated label (``LAB'') 
  & Close-up of the embossed ``LAB'' lettering on a matte laminated surface, highlighting fine spatial details and edges. \\
2 & Legumes 
  & Mixture of beans and small seeds with different shapes and sizes, providing heterogeneous organic textures. \\
3 & Salt 
  & Patch of coarse salt crystals forming a granular, high-frequency texture. \\
4 & Rice grains 
  & Dense layer of uncooked rice grains with subtle local variations due to grain orientation and packing. \\
5 & Corn kernels 
  & Compact arrangement of corn kernels with smooth curved surfaces and moderate specular reflections. \\
6 & Mixed cereals 
  & Heterogeneous mixture of breakfast cereals yielding complex spatial structure and organic spectral signatures. \\
7 & Rock surface 
  & Single rock fragment with visible roughness, edges, and shading variations. \\
8 & Coin 
  & Metallic coin on a dark background, emphasizing reflective properties and sharp contours. \\
\bottomrule
\end{tabular}
\end{table}

These 8 scenes are used consistently across all compared ordering strategies and sampling rates in the real-data experiments. For each scene, a full set of Hadamard measurements is first acquired and then numerically subsampled according to the different orderings, ensuring a fair comparison under identical optical and noise conditions.

\newpage

\section{Discussion}

The proposed  ordering of the Hadamard basis offers distinct advantages by integrating Deep Learning (DL) with the acquisition process while preserving the core benefits of the Hadamard transform. While our methodology roots its  ordering using the Hadamard matrix, its capability can be \textbf{adapted beyond this basis} to other types of generation patterns.

Firstly, the proposed method can complement methods employing end-to-end learned projection matrices such as~\cite{higham2018deep, wang2021single}. While learned projections provide high representational flexibility, our method's primary advantage is measurement efficiency, using a  selector for scene-dependent subset selection to significantly reduce the total number of patterns. A promising research direction is the integration where the adaptive ordering mechanism is trained to select the most relevant learned patterns from a compact, task-optimized set generated by a learned projection technique, thereby constructing and ordering a subset of learned patterns for acquisition, adapting the pattern set by relevance scene by scene.

Secondly, our method complements high-speed techniques using cyclic coded apertures~\cite{hahamovich2021single}. Cyclic masks excel at raw acquisition speed. Our advantage, conversely, is compressive efficiency, minimizing the required number of patterns. To expand our method's utility in this domain, the  selector can be trained to construct a subset of the cyclic Hadamard shifts and learn the optimal ordering for their acquisition. This directly combines the cyclic system's rapid per-pattern delivery with our substantial pattern reduction, compounding speed gains for high-frame-rate, compressive spectral imaging.

\clearpage 
\onecolumn 

\vfill

\end{document}